\documentclass{aa}             

\input epsf
\usepackage{longtable}

\begin{document}

\def\AD{AD\,Boo}
\def\HW{HW\,CMa}
\def\SW{SW\,CMa}
\def\V636{V636\,Cen}
\def\VZ{VZ\,Hya}
\def\WZ{WZ\,Oph}

\title{
Four-colour photometry of eclipsing binaries. XLI
\thanks{Based on observations carried out with the Str{\"o}mgren
Automatic Telescope (SAT) at ESO, La Silla, Chile}
}
\subtitle{
$uvby$ light curves for 
AD\,Bootis,            
HW\,Canis\,Majoris, \\ 
SW\,Canis\,Majoris,    
V636\,Centauri,        
VZ\,Hydrae, and        
WZ\,Ophiuchi.          
\thanks{
Tables~\ref{tab:adboo_tminp}--\ref{tab:wzoph_tmins}, and $6$--$13$
will be available on electronic form at the CDS via anonymous ftp 
to 130.79.128.5 or via http://cdsweb.u-strasbg.fr/Abstract.html.}
}
\author{
J.V. Clausen       \inst{1}
\and  L.P.R. Vaz \inst{2}
\and  J.M. Garc\'{\i}a \inst{3}
\and  A. Gim\'enez \inst{4,5}
\and  B.E. Helt \inst{1}
\and  E.H. Olsen \inst{1}
\and  J. Andersen \inst{1,6}
}
\offprints{J.V.~Clausen, \\ e-mail: jvc@astro.ku.dk}

\institute{
Niels Bohr Institute, Copenhagen University,
Juliane Maries Vej 30,
DK-2100 Copenhagen {\O}, Denmark
    \and
Depto.\ de F\'{\i}sica, ICEx, UFMG, C.P.\ 702 - 30.123-970 -
Belo Horizonte, MG, Brazil
    \and
Departamento de F{\'{\i}}sica, E.U.I.T. Industriales, UPM,
Ronda de Valencia 3, E-28012 Madrid, Spain
    \and
Laboratorio de Astrof{\'{\i}}sica Espacial y
F{\'{\i}}sica Fundamental, INTA, Apartado 50727,
      E-28080 Madrid, Spain
     \and
Research and Scientific Support Department,
ESA, ESTEC, NL-2200 AG Nordwijk, The Netherlands
     \and
Nordic Optical Telescope Scientific Association, Apartado 474, ES-38\,700
Santa Cruz de La Palma, Spain
}

\date{Received 28 February 2008 / Accepted 25 May 2008} 
 
\titlerunning{$uvby$ light curves}
\authorrunning{J.V. Clausen et al.}

\abstract
{Accurate mass, radius, and abundance determinations from
binaries provide important information on stellar evolution,
fundamental to central fields in modern astrophysics and cosmology.}
{Within the long-term Copenhagen Binary Project, we aim to obtain
high-quality light curves and standard photometry for
double-lined detached eclipsing binaries with late A, F, and G type 
main-sequence components, needed for the determination of accurate absolute 
dimensions and abundances, and for detailed comparisons with results from 
recent stellar evolutionary models.}
{Between March 1985 and July 2007, we carried out photometric observations
of \AD, \HW, \SW, \V636, \VZ, and \WZ\ at the Str{\"o}mgren Automatic Telescope 
at ESO, La Silla.}
{We obtained complete $uvby$ light curves, ephemerides, and standard 
$uvby\beta$ indices for all six systems.
For V636\,Cen and HW\,CMa, we present the first modern light curves, whereas
for AD\,Boo, SW\,CMa, VZ\,Hya, and WZ\,Oph, they are both
more accurate and more complete than earlier data.
Due to a high orbital eccentricity ($e = 0.50$), combined with a low orbital
inclination ($i = 84\fdg7$), only one eclipse, close to periastron, occurs
for HW\,CMa. 
For the two other eccentric systems, V636\,Cen ($e = 0.134$) and SW\,CMa 
($e = 0.316$), apsidal motion has been detected with periods of 
$5\,270 \pm 335$ and $14\,900 \pm 3\,600$ years, respectively.}
{}
\keywords{
Stars: evolution --
Stars: fundamental parameters --
Stars: binaries: close --
Stars: binaries: eclipsing --
Techniques: photometric}

\maketitle

\section{Introduction}
\label{sec:intro}

Detached, double-lined eclipsing binaries are well-known as the main source of
accurate (1--2\% or even better) data on stellar masses and radii.
During the last two-three decades, such results have been published for
about 50 sy\-stems within the main-sequence band (e.g. Andersen \cite{ja91}; 
Ribas et al. \cite{ir00a}; Southworth, private communication 
\footnote{{\scriptsize\tt http://www.astro.keele.ac.uk/$\sim$jkt/}}), 
but some mass intervals, especially around
and below 1 $M_{\sun}$, still need better coverage, 
as do parts of the main-sequence band at higher masses.  
A rich sample of accurate binary results is mandatory
for detailed tests of stellar models in general, as well as for many 
specific astrophy\-si\-cal investigations (Ribas \cite{ir04}, \cite{ir06b}). 
Remarkable recent examples are the
determination of the helium-to-metal enrichment ratio 
(Ribas et al. \cite{ir00a}) and investigations of the mass dependence of  
convective core overshooting (Ribas et al. \cite{ir00b}, Claret \cite{claret07}).

In this paper, we present new accurate and complete $uvby$ light curves
for six detached double-lined eclipsing binaries having late A, F, and
G-type main-sequence components. In addition, new spectroscopic observations
for radial velocities and abundance analyses have been obtained, and 
detailed analyses for all the systems will be published se\-pa\-ra\-tely
(Clausen et al. \cite{avw08}, Clausen et al. in prep., Torres et al. in prep.).

When we decided to observe AD\,Boo (F6+G0) and SW\,CMa (A8), 
they were lacking modern photometric and spectroscopic observations.
This was independently noticed by Lacy (\cite{lacy97a}, \cite{lacy97b}) 
who studied both sy\-stems, and by Popper (\cite{dmp98a}) who obtained
spectroscopic orbits for AD\,Boo, leading to masses for the components
which differ somewhat from those by Lacy. 
For both sy\-stems, a new analysis based on new light curve and radial velocity 
observations, supplemented with a spectroscopic abundance analysis, 
is desirable.

HW\,CMa (A8) is only about $2 \arcmin$ distant from SW\,CMa 
and was discovered to be a double-lined spectroscopic binary du\-ring the 
radial velocity observations of SW\,CMa. It was later found to be eclip\-sing 
(Liu et al. \cite{liu92}), and we decided to obtain complete light curves 
and spectroscopic orbits. 

V636\,Cen (F8/G0V) was selected as part of a large-scale study of new systems 
with solar-type components (Clausen et al. \cite{jvcetal01}).
We want e.g. to investigate a serious dilemma that appears to be present in the
compa\-ri\-son of predictions from current stellar models with fundamental
properties of known $0.7$--$1.1$ $M_{\sun}$ eclipsing binaries
(Popper~\cite{dmp97}, Clausen et al.~\cite{granada99b}).
Our results for V636\,Cen support the presence of a real discrepancy -  
current models, adopting a fixed mixing length parameter calibrated
on the Sun for envelope convection, do not predict identical ages for 
the two components (Larsen \cite{al98}, Clausen et al. in prep.). 
A similar case, V1061\,Cyg, was presented by Torres et al. 
(\cite{wt06}). See also Ribas (\cite{ir06a}) for eclipsing binaries with 
K and M type components and Morales et al. (\cite{mrj07}) for single
active late-K and M stars.

VZ\,Hya (F5) and WZ\,Oph (F8) are well known systems included by Popper 
(\cite{dmp80}) in his critical review of stellar masses, but
excluded by Andersen (\cite{ja91}), since the available 
results did not fulfill his 2\% criteria for masses and radii. 
We therefore decided to provide the new observations needed for the
determination of more accurate dimensions.

In the following, we first give a general description of the observations,
the basic photometric reduction, and the light curve formation 
(Sect. \ref{sec:obs}), and we then present the results for the 
individual systems (Sect. \ref{sec:adboo} -- \ref{sec:wzoph}).
Throughout the paper, the component eclipsed at the deeper eclipse at phase 
0.0 is referred to as the primary, and the other as the secondary component.

\section{Observations and photometric reduction}
\label{sec:obs}

The differential $uvby$ light curve observations were obtained with 
the Str{\"o}mgren Automatic Telescope (SAT) at ESO, La Silla 
and its 6-channel $uvby\beta$ photometer during se\-ve\-ral
campaigns between March 1985 and March 2002. For SW\,CMa and V\,636\,Cen, 
more eclipse observations were carried out between January 2001 and July 2007  
to follow the apsidal motion of their orbits.
In addition, on many nights we observed the binaries and the comparison 
stars together with a large sample of $uvby$ and $\beta$ standard stars 
in order to obtain homogeneous standard $uvby\beta$ indices, primarily for
the determination of effective temperatures and interstellar reddening.
Details on the photometer and the (semi)automatic 
mode of the telescope are given by Olsen (\cite{olsen93}, \cite{olsen94b}).

\subsection{Light curve observations}
\label{sec:lcobs}

\begin{table}
\caption[]{\label{tab:comparisons_dm}
rms errors of the magnitude differences (instrumental system) between 
the comparison stars in units of 0.1 mmag.
N is the total number of magnitude differences.
}
\begin{center}
\begin{tabular}{lrrrrr} \hline
\hline\noalign{\smallskip}
Objects                    &  N & $y$ & $b$ & $v$ & $u$ \\ 
\noalign{\smallskip}
\hline
\noalign{\smallskip}
AD\,Boo                    &     &     &     &     \\
HD128369, HD129430 (C1,C2) & 196 &  47 &  45 &  46 &  68  \\
HD128369, HD128185 (C1,C3) & 164 &  58 &  57 &  62 &  85 \\
HD129430, HD128185 (C2,C3) & 141 &  67 &  59 &  66 &  81 \\
\hline
\noalign{\smallskip}
HW\,CMa \& SW\,CMa         &     &     &     &     &     \\
HD56341, HD53123 (C1,C2)   & 536 &  54 &  45 &  48 &  54  \\
\hline
\noalign{\smallskip}
V636\,Cen                  &     &     &     &     &      \\
HD124829, HD125444 (C1,C2) & 388 &  51 &  52 &  54 &  65 \\
\hline
\noalign{\smallskip}
VZ\,Hya                    &     &     &     &     &     \\
HD72528,  HD71615  (C1,C2) & 328 &  37 &  35 &  37 &  59 \\
HD72528,  HD72782  (C1,C3) & 333 &  43 &  36 &  41 &  61 \\
HD71615,  HD72782  (C2,C3) & 335 &  43 &  33 &  36 &  57 \\
\hline
\noalign{\smallskip}
WZ\,Oph                    &     &     &     &     &     \\
HD155193, HD154931 (C1,C2) & 288 &  40 &  38 &  39 &  61 \\
\hline
\end{tabular}
\end{center}
\end{table}

\begin{table*}
\caption[]{\label{tab:comparisons_inf}
New photometric data for the eclipsing binaries and the comparison stars.
For the eclipsing binaries, the $uvby\beta$ information is the mean values outside eclipses,
see also Fig~\ref{fig:beta}.
N is the total number of observations used to form the mean values, and
$\sigma$ is the rms error (per observation) in  mmag. See also Table~\ref{tab:std_publ}.
}
\begin{center}
\begin{tabular}{llrrrrrrrrrrrr} \hline
\hline\noalign{\smallskip}
Object&Sp. Type&$V$&$\sigma$&$b-y$&$\sigma$&$m_1$&$\sigma$&$c_1$ &$\sigma$&N($uvby$)&$\beta$&$\sigma$&N($\beta$)\\
\noalign{\smallskip}
\hline
\noalign{\smallskip}
%
AD\,Boo  & F6+G0               & 9.379  & 7& 0.322 & 4& 0.186 &10& 0.431 &14& 14 & 2.647   & 8 &16\\ 
HD128369 & F5                  & 7.959  & 8& 0.281 & 3& 0.149 & 8& 0.455 &11& 17 & 2.640   & 8 &15\\
HD129430 & G8III-IV            & 6.402  & 5& 0.581 & 2& 0.311 & 6& 0.385 & 9&  5 & 2.573   & 3 & 3\\
HD128185 & F8V                 & 7.925  &11& 0.373 & 5& 0.183 &10& 0.402 &12& 13 & 2.606   & 8 &14\\
\hline
\noalign{\smallskip}
HW\,CMa  & A8                   & 9.190  & 6& 0.128 & 3& 0.228 &10& 0.839 &14&  6 & 2.853   &10 &95\\
SW\, CMa & A8                   & 9.149  & 9& 0.094 & 8& 0.210 & 9& 1.013 & 7& 12 & 2.865   &10 &104\\ 
HD56341 & A0V                  & 6.358  & 8&-0.001 & 2& 0.127 & 5& 1.098 & 7& 17 & 2.861   & 6 &53\\
HD53123 & B9V                  & 7.156  &13&-0.020 & 5& 0.112 & 7& 0.878 & 9& 10 & 2.805   & 8 &51\\
HD55271 & B5V                  & 6.950  &14& 0.005 & 2& 0.060 & 8& 0.290 & 6&  2 & 2.520 &10  & 9\\
\hline
\noalign{\smallskip}
V636\,Cen& F8/G0V              & 8.704  &11& 0.410 & 4& 0.205 & 7& 0.285 &18& 16 & 2.594   &12 &57\\
HD124829 & F2                  & 8.466  & 6& 0.266 & 4& 0.143 & 6& 0.516 & 6& 16 & 2.682   & 6 &16\\
HD125444 & A6V                 & 7.557  &12& 0.111 & 4& 0.176 & 7& 0.918 & 5& 10 & 2.814   & 9 &34\\
\hline
\noalign{\smallskip}
VZ\,Hya  & F5+F6               & 8.953  & 4& 0.300 & 4& 0.166 & 7& 0.389 & 9&  9 & 2.640   & 9 &83\\ 
HD72528  & F7V                 & 7.330  & 3& 0.337 & 2& 0.177 & 3& 0.388 & 6&  8 & 2.628   & 8 &33\\
HD71615  & F0                  & 7.946  & 3& 0.233 & 4& 0.176 & 7& 0.516 & 8&  9 & 2.692   & 8 &34\\
HD72782  & A2                  & 7.641  &10& 0.109 & 2& 0.183 & 3& 0.947 &11&  5 & 2.831   & 7 &61\\ 
\hline
\noalign{\smallskip}
WZ\,Oph  & F8                   & 9.096  & 7& 0.366 & 5& 0.145 & 7& 0.375 & 5& 11 & 2.627   & 8 &11\\
HD155193 & F8IV                & 7.007  & 7& 0.338 & 4& 0.185 & 5& 0.405 & 6& 12 & 2.623   & 7 &12\\
HD154931 & G0                  & 7.258  & 7& 0.384 & 4& 0.189 & 6& 0.361 & 4& 12 & 2.600   & 8 &13\\
\hline
\end{tabular}
\end{center}
\end{table*}
\begin{table*}
\caption[]{\label{tab:std_publ}
Published photometric data for the eclipsing binaries and the comparison stars.
N is the total number of observations used to form the mean values, and
$\sigma$ is the rms error (per observation) in  mmag.
References are:
HH75 =  Hilditch \& Hill (\cite{hh75}),        
J96  = Jordi et al. (\cite{jordi96}),          
O83 = Olsen (\cite{olsen83}),                  
O93 = Olsen (\cite{olsen93}),                  
O94 = Olsen (\cite{olsen94b}),                 
OP83 = Olsen \& Perry (\cite{olsenperry84}),   
P69 = Perry (\cite{perry69}),                  
WK83 = Wolf \& Kern (\cite{wk83})              
L02 = Lacy (\cite{lacy02}).                    
The two WK83 observations of VZ Hya are done near phase 0.00.
Two WK83 observations of SW CMa inside eclipses have not been included.
Two HH75 observations of WZ Oph inside eclipses have not been included.
See also Table~\ref{tab:comparisons_inf}.
}
\begin{center}
\begin{tabular}{llrrrrrrrrrrrr} \hline
\hline\noalign{\smallskip}
Object&Reference&$V$&$\sigma$&$b-y$&$\sigma$&$m_1$&$\sigma$&$c_1$ &$\sigma$&N($uvby$)&$\beta$&$\sigma$&N($\beta$)\\
\noalign{\smallskip}
\hline
\noalign{\smallskip}
AD\,Boo &  HH75      &  9.370  &  & 0.348 &  & 0.153 &  & 0.419 &  &  1 &         &   &  \\
        &  L02       &  9.399  &  & 0.335 &  & 0.161 &  & 0.446 &  &  1 & 2.645   &   & 1\\
HD128369&  O83       &  7.965  & 3& 0.298 & 3& 0.122 & 5& 0.453 & 5&  5 & 2.643   & 2 & 3\\
HD129430&  O93       &  6.401  & 4& 0.590 & 2& 0.293 & 2& 0.413 & 3&  1 &         &   &  \\
HD128185&  O83       &  7.925  & 4& 0.387 & 6& 0.163 & 2& 0.420 & 3&  4 & 2.604   & 8 & 3\\
\hline
\noalign{\smallskip}
SW\,CMa &  L02       &  9.121  &  & 0.108 &  & 0.221 &  & 1.026 &  &  1 & 2.860   &   & 1\\
        &  WK83      &  9.15   &  & 0.095 &  & 0.209 &  & 1.036 &  &  1 & 2.863   &   & 1\\ 
\hline
\noalign{\smallskip}
V636\,Cen& O83,OP84  &  8.713  & 6& 0.407 & 5& 0.220 & 9& 0.259 & 7&  1 & 2.601   & 7 & 1\\
        &  O85       &  8.706  & 7& 0.426 & 3& 0.197 & 5& 0.276 & 8&  3 &         &   &  \\
        &  O93       &  8.720  & 2& 0.414 & 3& 0.205 & 6& 0.284 & 8&  3 &         &   &  \\
\hline
\noalign{\smallskip}
VZ\,Hya &  WK83      &  9.660  &30& 0.333 & 1& 0.146 & 1& 0.368 & 4&  2 & 2.624   & 7 & 2\\
HD72528 &  O83,OP84  &  7.335  & 6& 0.328 & 4& 0.181 & 6& 0.381 & 6&  1 & 2.617   & 7 & 1 \\
HD71615 &  O83,OP84  &  7.949  & 5& 0.249 & 4& 0.141 & 6& 0.520 & 7&  1 & 2.685   & 7 & 1 \\
HD72782 &  O83       &  7.648  & 6& 0.103 & 5& 0.193 & 9& 0.947 &15&  2 &         &   &  \\
        &  J96       &         &  &       &  &       &  &       &  &    & 2.840   & 9 & 2\\ 
\hline
\noalign{\smallskip}
WZ\,Oph &  HH75      &  9.095  & 7& 0.377 &32& 0.119 &42& 0.367 &10&  2 &         &   &  \\
HD155193&  O83       &  7.010  & 5& 0.352 & 3& 0.168 & 4& 0.419 & 4&  1 &         &   &  \\
        &  O94       &  7.020  & 4& 0.350 & 3& 0.163 & 4& 0.429 & 4&  1 &         &   &  \\
        &  P69       &  7.020  &  & 0.348 &  & 0.168 &  & 0.418 &  &  2 &         &   &  \\
HD154931&  O83       &  7.263  & 9& 0.385 & 7& 0.192 & 0& 0.372 & 7&  2 &         &   &  \\
        &  O94       &  7.251  & 4& 0.396 & 3& 0.167 & 4& 0.378 & 5&  1 &         &   &  \\
        &  P69       &  7.250  &  & 0.392 &  & 0.179 &  & 0.365 &  &  2 &         &   &  \\
\hline
\end{tabular}
\end{center}
\end{table*}

The differential $uvby$ observations were, depending on the orbital phase of 
the binary, carried out either continuously over several hours or as shorter 
series.
Automatic centering within a circular diaphragm of 
$13\arcsec$ (HW\,CMa, part of light curves; SW\,CMa, light curves; V636\,Cen, 
light curves) or 
$17\arcsec$ (AD\,Boo, light curves; HW\,CMa, part of light curves; 
SW\,CMa, additional eclipse observations;
V636\,Cen, additional eclipse observations; VZ\,Hya, light curves; 
WZ\,Oph, light curves) 
diameter was used throughout.
Nearly all observations were performed at an airmass less than 2.1, and the few 
at higher airmass passed additional quality checks.

\begin{figure*}
\epsfxsize=185mm
\epsfbox{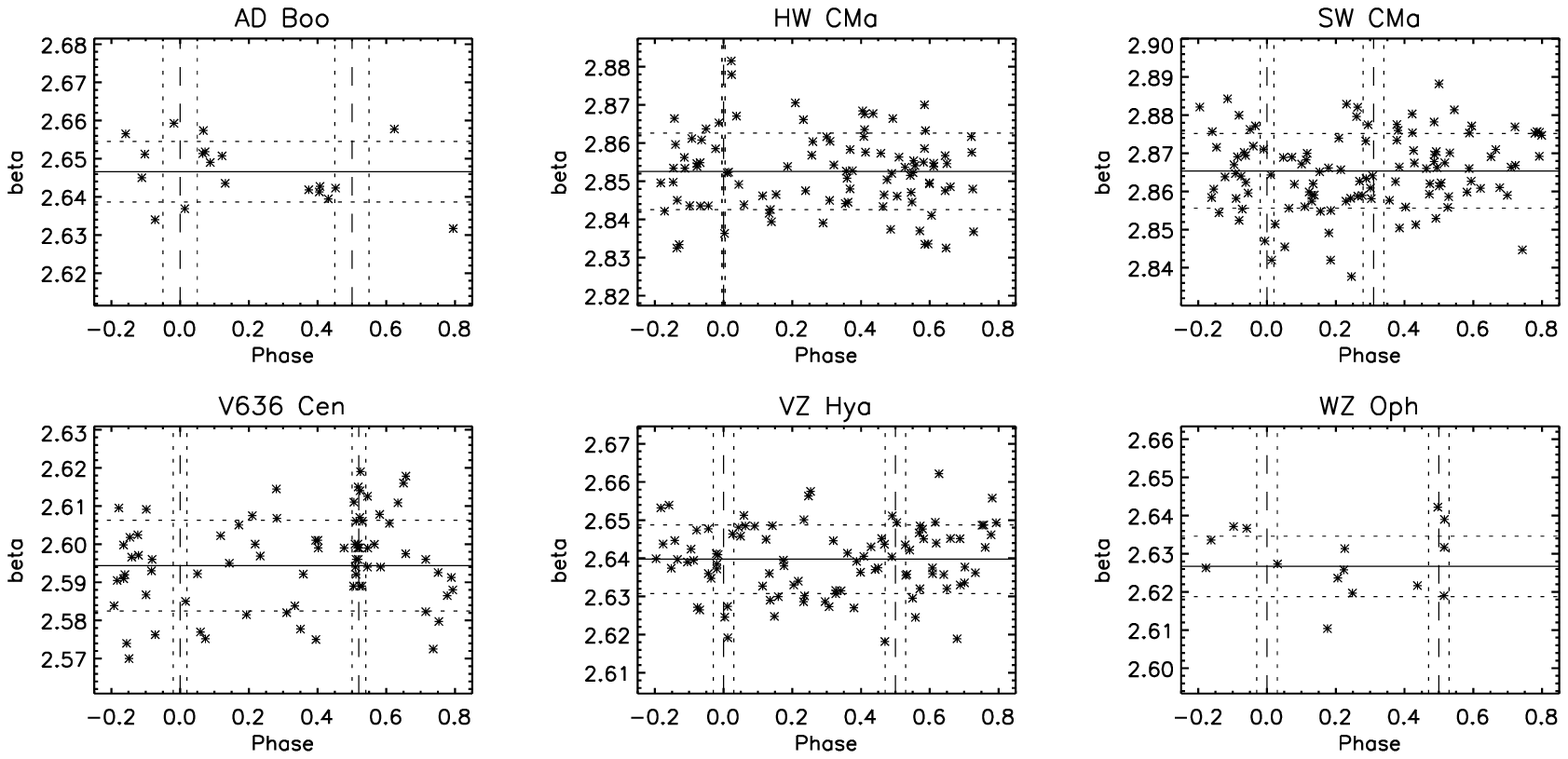}
\caption[]{\label{fig:beta}
Phase distribution of the $\beta$ observations, 
see Table~\ref{tab:comparisons_inf}.
The horizontal lines represent the mean value of $\beta$ outside 
eclipses ($solid$) and the rms error of one observation ($dotted$). 
The vertical lines show the phase intervals of the eclipses ($dashed$, 
central eclipses) and their duration ($dotted$).
}
\end{figure*}

Two or three comparison (C) stars, preferably matching the spectral type of 
the binary and positioned on the sky within a few degrees from it, were 
adopted for each candidate. 
They were in general selected from systematic searches in the extensive 
$uvby\beta$ catalogues published by Olsen (\cite{olsen83}, \cite{olsen88}, 
\cite{olsen93}, \cite{olsen94a}, \cite{olsen94b}), and unpublished $\beta$ 
photometry (cf.  Olsen \cite{olsen94b}), avoiding known and su\-spected 
variables, as well as most visual binaries.

The observations were carried out in C1-binary-C2-binary-C3-binary-C1-etc. or 
C1-binary-C2-binary-C1-etc. sequen\-ces, allowing accurate magnitude 
differences to be formed, even if one of the comparisons were later found 
to be variable. 
In the case of HW\,CMa and SW\,CMa, which are close on the sky and of similar 
spectral types, we often used C1-SW\,CMa-C2-HW\,CMa-C1-etc.
An observation consisted in general of three individual integrations, each with
an integration time of 10--60 s, and sky measurements were taken at least
once per sequence, normally at a fixed position close to the binary. 
The resulting rms contributions from photon statistics, including the
sky contributions, were in general kept at 5 mmag or lower.
The Heliocentric Julian Date (HJD) given for an observation refers to the
midpoint of the time interval co\-ve\-red by the integrations.

\subsection{Photometric reduction and formation of light curves}
\label{sec:red}

Linear extinction coefficients were determined individually for each night
from the observations of comparison stars and other constant stars. Whenever
appropriate, linear or quadratic corrections for drift during the night,
caused by changes in the sky transparency and/or the influence of temperature 
variations on the uncooled photomultipliers, were also applied.

The $uvby$ instrumental system of the SAT, where the same spectrometer, 
including its filters, most of the photomultipliers, and the photon counting 
system have been used since 1985, has proved to be very
stable (e.g. Olsen et al., in prep.). Therefore, transformation of the light 
curve observations, e.g. to the standard $uvby$ system, is not needed, and the
light curves are generated in the instrumental sy\-stem.

Differential magnitudes were formed for each binary observation 
from the two comparison star observations closest in time.
All comparison star observations were used with C2 and C3, if observed, 
first shifted to the level of C1. A careful check of the constancy of all 
two/three stars was performed. For each binary, typical rms errors per light 
curve point are close to those listed in Table~\ref{tab:comparisons_dm} 
for the magnitude differences between the corresponding compa\-ri\-son stars.

Ephemerides were calculated from published and new
times of minima, listed in Tables~\ref{tab:adboo_tminp}--\ref{tab:wzoph_tmins}.
For most of the systems, our literature searches were significantly 
supplemented by the compilations by Kreiner (private communication); see 
Kreiner et al. (\cite{kreiner01}) \footnote{http://www.as.ap.krakow.pl/ephem}
for further details.
The new times of minima have been calculated using the method of
Kwee and van Woerden (\cite{kvw56}), except for a few cases
which could only be determined by 2nd order polynomial fits to the (few) 
observations. 

\begin{figure*}
\epsfxsize=185mm
\epsfbox{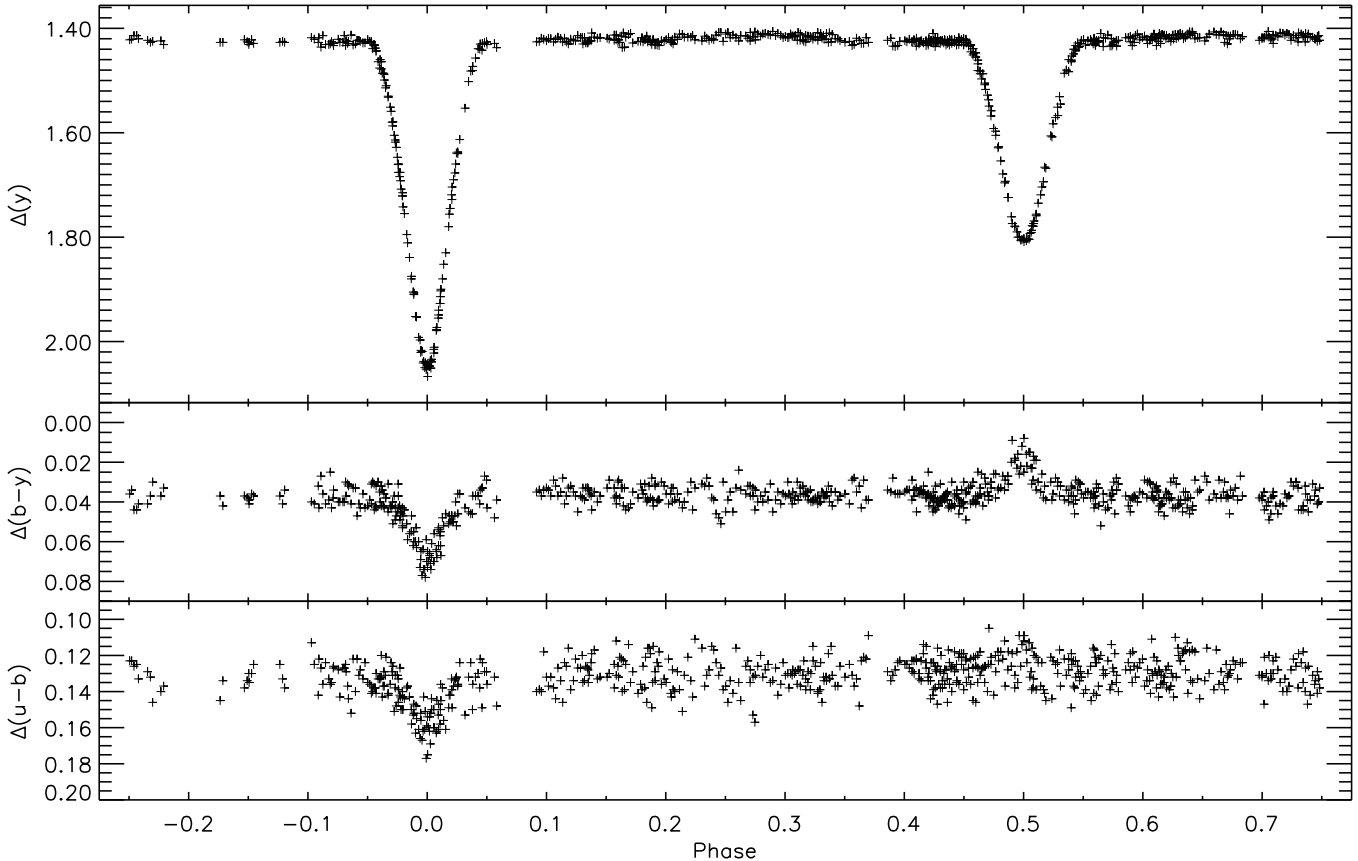}
\caption[]{\label{fig:adboo}
$y$ light curve and $b-y$ and $u-b$ colour curves (instrumental system)
for AD\,Boo. 
}
\end{figure*}

The individual light curves are shown in 
Figs.~\ref{fig:adboo}, \ref{fig:hwcma}, \ref{fig:swcma}, \ref{fig:v636cen},
\ref{fig:vzhya}, and \ref{fig:wzoph}, and are presented in Tables~$6$--$11$.
The additional eclipse observations of \SW\ and \V636\ are presented in
Tables~12 and 13.
These eight tables will only be made available in electronic form.

\subsection{Standard $uvby\beta$ photometry}
\label{sec:std}

New standard $uvby\beta$ indices for the six eclipsing binaries 
outside eclipses and for all comparison stars are listed in 
Table~\ref{tab:comparisons_inf}, 
and the phase distributions of the $\beta$ observations are shown in 
Fig.~\ref{fig:beta}.

Part of the standard photometry was obtained at SAT on several nights 
in 1987--1993,
where the binaries and the comparison stars were observed together with
a large sample of $uvby$ and $\beta$ standard stars. 
The basic reduction and transformation to the standard system was done as 
described by Olsen (e.g. \cite{olsen93}, \cite{olsen94b}).
Furthermore, some of the binaries and their comparisons were also included 
in the long term search for new solar-type binaries, carried out at SAT 
since 1994, and additional standard photometry from many nights has been 
obtained from this project; see Olsen et al. (in prep.) for details. 

For comparison, we have listed published $uvby\beta$ indices for some 
of the binaries and their comparison stars in Table~\ref{tab:std_publ}. 
Individual differences larger than the quoted errors occur; we have used
the new homogeneous results in Table~\ref{tab:comparisons_inf} for the
binary analyses.

\section{\object{AD\,Boo}} 
\label{sec:adboo}

AD\, Boo = BD +25\degr 2800 = BV 135 was discovered by Strohmeier et al. 
(\cite{sm56}) to be an eclipsing binary, and times of minima, a linear 
ephemeris, and a photographic light curve were published by Strohmeier 
et al. (\cite{sm63}; orbital period half the correct value).
A large number of times of minima of AD\, Boo has later been observed 
(see Tables~\ref{tab:adboo_tminp} and \ref{tab:adboo_tmins}), 
and an improved ephemeris and a visual light curve was published by 
van Buren (\cite{vb74}; orbital period half the correct value). 
Koch (\cite{koch74}) presented blue CN-absorption measurements, and
Zhai et al. (\cite{zhai82}, \cite{zhai83}) obtained photoelectric $B,V$ light
curves and derived the true orbital period of about $2\fd 07$.

Lacy (\cite{lacy85}) discovered AD\,Boo to be a double-lined spectroscopic
binary, and later (Lacy \cite{lacy97a}) he presented absolute dimensions
based on spectroscopic elements from 23 KPNO CCD spectra and photometric
elements derived from EBOP (Popper \& Etzel \cite{dmpetzel81}) re-analyses of 
the light curves (normal points) by Zhai et al.  (\cite{zhai82}). 
AD\,Boo was included by Popper in his extensive study of F--K binaries 
(e.g. Popper \cite{dmp93}, Popper \& Jeong \cite{dmpj94}, Popper \cite{dmp96}),
and he obtained a very accurate spectroscopic orbit based on 31 Lick CCD 
echelle spectra (Popper \cite{dmp98a}). 
He also pointed out that the components of AD\,Boo have the
largest differences in mass and radius among well-studied systems in
its mass range (1.4 and 1.2 $M_{\sun}$, 1.6 and 1.2 $R_{\sun}$). 
Hence, it may provide stringent tests of stellar mo\-dels.

\subsection{$uvby$ light curves}
\label{sec:adboo_lc}

Complete $uvby$ light curves containing 
652 points in each colour were observed on 42 nights during five pe\-ri\-ods 
between March 1988 and March 1992 (JD2447234--JD2448695).
\object{HD128185}, \object{HD128369}, and \object{HD129430} = HR5483
were selected as comparison stars; 
see Tables~\ref{tab:comparisons_dm}, \ref{tab:comparisons_inf}, and
\ref{tab:std_publ} for further information.
The last two stars were found to be constant within the observational accuracy 
of a few mmag during our observations, whereas signs of a slight long term
variability at the level of about $\pm 5$ mmag were noticed for 
HD128185, and it was consequently decided to reject it for the calculations of 
the final light curves. We have not studied the possible variations of HD128185
further; our photometric data for it are available on request.

The light curves of AD\, Boo are shown in Fig.~\ref{fig:adboo}.
The eclipses have been covered several times and most out--of--eclipse phases 
at least twice.
The accuracy per point is about 0.005 mag ($vby$) and 0.007 mag ($u$), but
comparison of the data from the five periods reveals
that at some phases, systematic differences in the light level exist,
of the order of 0.01--0.03 mag, and increasing from $y$ to $u$. 
Especially the observations during eclipses from 1989 are systematically 
fainter.
The scatter is probably caused by slight activity
of the cooler secondary component, since a weak emission feature
in the Ca~II H and K lines is seen at the position of the secondary
component on high-resolution spectra of AD\,Boo (Clausen et al., \cite{avw08}).
Zhai et al. (\cite{zhai82}) also noticed scatter in their $B$ and $V$ light
curves, which were obtained from April to June 1981, but at a much
higher level of about 0.075 mag, compared to their obser\-ving error of
about 0.02 mag.
 
The new $uvby$ light curves confirm that AD\,Boo consists of two 
well-detached and rather different stars in a circular orbit, and 
our photometric analysis reveals that the secondary eclipse is in fact total.

\begin{figure*}
\epsfxsize=185mm
\epsfbox{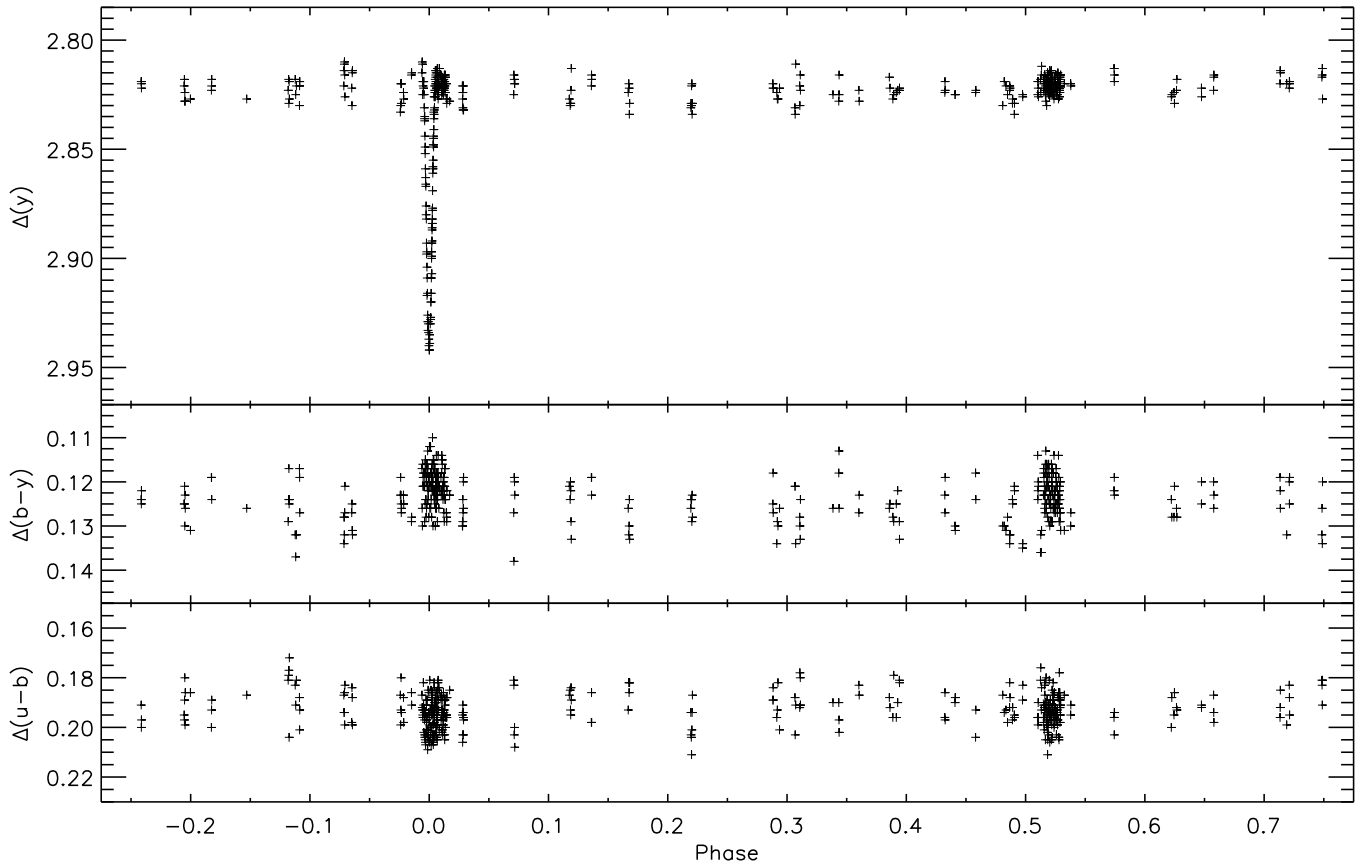}
\caption[]{\label{fig:hwcma}
$y$ light curve and $b-y$ and $u-b$ colour curves (instrumental system)
for HW\,CMa.
}
\end{figure*}

\subsection{Ephemeris}
\label{sec:adboo_eph}

Four times of primary minimum and two of secondary, derived 
from the $uvby$ observations, are listed in Tables~\ref{tab:adboo_tminp} and
\ref{tab:adboo_tmins} together with 130 published times.

The linear ephemeris given in Eq.~\ref{eq:adboo_eph} was computed by 
least squares based on all
available visual, photographic, and photoelectric/CCD times of
eclipses, assuming a cir\-cu\-lar orbit. 
\begin{equation}
\label{eq:adboo_eph}
\begin{tabular}{r r c r r}
{\rm Min \, I} =  & 2449311.43169 & + & $2\fd 06880704$ &$\times\; E$ \\
                  &      $\pm  22$&   &       $\pm  17$ &             \\
\end{tabular}
\end{equation}
\noindent
Weights were assigned to the observations based on the
individual uncertainties, when available. Typically the visual and
photographic minima have no published uncertainties, so those were
determined iteratively based on the residuals from the fit, in such a
way as to achieve a reduced $\chi^2$ of unity for each type of
observation.  Similarly, scale factors were determined for the
photoelectric/CCD measurements so as to yield a reduced $\chi^2$ of
unity, separately for the primary and secondary minima, since they can
have different precisions due to the different depth of the eclipses.

Separate least squares fits to all primary and secondary times
yield identical periods within errors, and  again within errors the same 
ephemeris is obtained from the subset of photoelectric and CCD data. 
The method of Lafler \& Kinman (\cite{lk65}) applied to the $uvby$
eclipse observations also confirms the period.  

Zhai et al. (\cite{zhai83}) derived a 
slightly longer period of $2 \fd 0688112 \pm 0.0000002$ from their times 
of minima and those listed by van Buren (\cite{vb74}), and their ephemeris 
was used by Lacy (\cite{lacy97a}) and Popper (\cite{dmp98a}).
We see, however, no clear signs of period changes from the data
listed in Tables \ref{tab:adboo_tminp} and \ref{tab:adboo_tmins}.

\section{\object{HW\,CMa}} 
\label{sec:hwcma}

HW\,CMa = HD54549 was discovered by Liu et al. (\cite{liu92}) 
to be a double-lined eclipsing binary in a highly eccentric
orbit ($P = 21\fd1$, $e$ = 0.5). Minimum masses of 1.74 and 1.80 $M_{\sun}$ 
were determined, and a shallow eclipse (about 0.13 mag deep) lasting for
about 5 hours (0.010 in phase) was detected at conjunction, which occurs close 
to periastron.  The star eclipsed is the less massive one.
At apastron, the separation between the components is about three times
larger, and no eclipse is seen around the predicted phase (0.517).
This is due to the high orbital eccentricity, combined with a low
orbital inclination ($i = 84\fdg7$). 
Houk \& Smith-Moore (\cite{hsm88}) classify HW\,CMa as A1\,III, but CfA 
spectra and $uvby$ photometry reveal that the components are almost identical
unevolved main-sequence stars near A8. 

\subsection{$uvby$ light curves}
\label{sec:hwcma_lc}

The $uvby$ light curves, which  contain 
414 points in each colour, were observed on 53 nights between February 1989 
and March 2002 (JD2447576--JD2452335). The observations by
Liu et al. (\cite{liu92}), with the same system, have been included.
HW\,CMa is only about $2\arcmin$ distant from SW\,CMa and of similar
spectral type. Therefore, we have used the same comparison stars, 
HD56341 = HR2755 and HD53123, and as mentioned in Sect.~\ref{sec:lcobs}
HW\,CMa was often observed in series which also included SW\,CMa.
For the October 2001--March 2002 period, a third star 
HD55344 was added as a further check of the
compari\-son stars, which were found to be constant within the observational 
accuracy; see Tables~\ref{tab:comparisons_dm}, \ref{tab:comparisons_inf}, 
and \ref{tab:std_publ} for further information.

The light curves are shown in Fig.~\ref{fig:hwcma}. The eclipse at phase
0.0 was observed on three nights, and the expected interval near phase
0.517 for the other eclipse, which should last significantly longer, 
was covered closely on one night with a few additional observations 
from other nights. 
As mentioned above, no sign of an eclipse is, however, seen. 
As expected, the light curves are essentially constant outside eclipse, 
and no significant colour changes are seen during the eclipse, where 
$\sim$20\% of the light of the primary component is blocked.

\begin{figure*}
\epsfxsize=185mm
\epsfbox{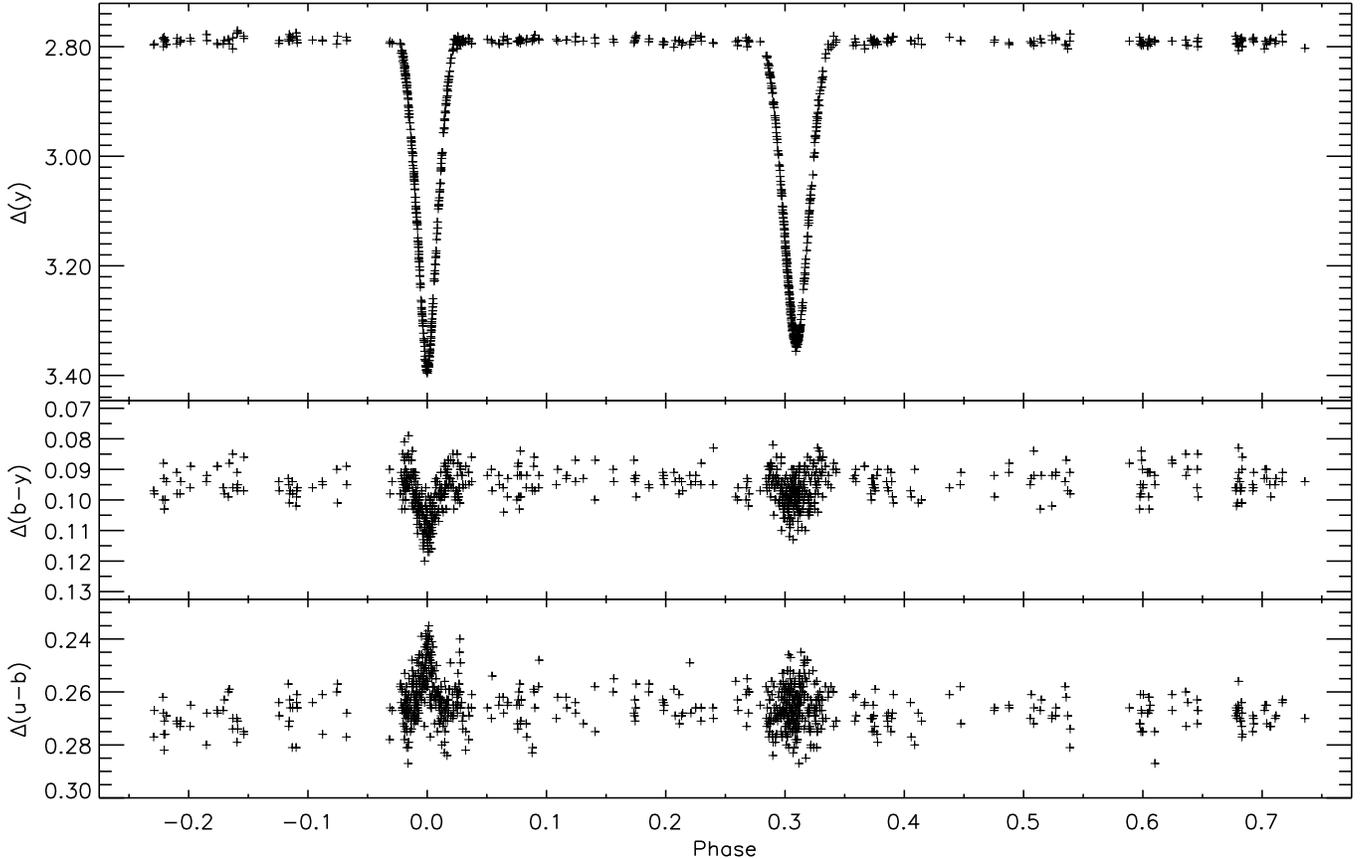}
\caption[]{\label{fig:swcma}
$y$ light curve and $b-y$ and $u-b$ colour curves (instrumental system)
for SW\,CMa.
}
\end{figure*}

\subsection{Ephemeris}
\label{sec:hwcma_eph}

Two minimum times, separated by 170 epochs, have been determined
from the $uvby$ observations; see Table~\ref{tab:hwcma_tmin}.
This limited material yields the following linear ephemeris for
HW\,CMa:

\begin{equation}
\label{eq:hwcma_eph}
\begin{tabular}{r r c r r}
{\rm Min \, I} =  & 2452279.6787 & + & $21\fd 1178329$ &$\times\; E$ \\
                  &      $\pm  4$&   &       $\pm  33$ &             \\
\end{tabular}
\end{equation}

\noindent
For comparison, Liu et al. (\cite{liu92}) derived a period of
$21\fd 1178 \pm 0\fd 0005$, and the spectroscopic orbit based on
the CfA velocities, which cover 178 cycles, gives 
$21\fd 11790 \pm 0\fd 00025$.
Since the orbit is highly eccentric, apsidal motion is expected to
be present, although with a long period due to the large separation
between the components and the correspondingly small relative radii. 
The orbital parameters and estimated stellar dimensions yield a 
theoretically predicted period of about 90\,000 yr. 

\section{\object{SW\,CMa}} 
\label{sec:swcma}

SW\,CMa = HD54520 = HIP34431 was discovered by Hoffmeister (\cite{h32}) to
be an eclipsing binary, and Florja (\cite{florja37}) observed it visually
and determined its ephemeris (Min =  $2426709.324 + 10 \fd 092 \times E$). 
The period is close to the correct value, but today we know that the epoch 
corresponds to secondary eclipse. Floja's visual light curve shows only one 
eclipse, and he therefore speculates whether one of the only two times where 
Hoffmeister found a minimum, which does not fit the ephemeris, is wrong.
Again, today we know that this observation is correct and corresponds 
to the primary minimum.

Struve (\cite{struve45}) found SW\,CMa to be a double-lined spectroscopic
binary with $\approx 2 M_{\sun}$ components in a very eccentric orbit 
($e = 0.5$; significantly higher than the correct value of about 0.3).
Later, a photographic light curve and times of minima, based on the old
Sonneberg plates used by Hoffmeister, were presented by
Wenzel (\cite{wenzel52}), and several additional times of minima are
available in the literature (see Table \ref{tab:swcma_tmin}).

Popper (\cite{dmp65}) mentioned SW\,CMa as one of several southern late B 
to early A systems that should be stu\-died, and, using modern detectors, 
Lacy (\cite{lacy84}) found narrow, deep lines of nearly equal strength for 
the components, and encouraged photoelectric observers to obtain accurate 
light curves.
$uvby\beta$ indices have been published by Wolf \& Kern (\cite{wk83}),
and $UBV$ indices by Lacy (\cite{lacy92}). 
SW\,CMa was included by Gim\'enez (\cite{agc94}) in his list of eccentric 
binaries that should be monitored photo\-me\-tri\-cal\-ly in order to study 
apsidal motion, and finally Lacy (\cite{lacy97b}) determined absolute 
dimensions from analyses of $UBV$ light curves observed at CTIO 1993--95 and 
coud\'e spectra observed at McDonald and KPNO 1982--89. 
His analysis reveals that SW\,CMa consists of two somewhat different 
components (2.2 and 2.0 $M_{\sun}$, 3.0 and 2.5 $R_{\sun}$)
which have evolved to the upper part of the main-sequence band. 

\subsection{$uvby$ light curves}
\label{sec:swcma_lc}

Complete $uvby$ light curves containing 820 points in each colour were 
observed on 89 nights during five periods between February 1987 and
March 1992 (JD2446825--JD2448689).
As mentioned in Sect.~\ref{sec:lcobs},
SW\,CMa was often observed in series which also included HW\,CMa.
For the 1987 observations, we selected \object{HD55271} 
and \object{HD56341}  = HR2755 as comparison stars, and in
addition \object{HD53123} was observed frequently on each night. 
None of the stars shows any sign of va\-ria\-bi\-li\-ty within the 
observational accuracy of a few mmag; see Tables~\ref{tab:comparisons_dm} and 
\ref{tab:comparisons_inf}. However, HD55271 = ADS5863A = WDS 07113-2148A
(A-BC: $\rho = 13 \farcs 6$, $\Delta m = 1.3$, Worley \& Douglass \cite{wds})
is a member of a double star, and automatic centering of it proved to be less
precise than desirable due to the companion. HD55271 was therefore
rejected as comparison for the 1988--92 observations and replaced by HD53123. 
We have used the HD56341 and HD53123 observations for calculation of the light 
curves.

The new $uvby$ light curves, which are shown in Fig.~\ref{fig:swcma}, 
contain about twice the number of points per band as the $UBV$ light curves 
by Lacy (\cite{lacy97b}), and they are of at least the same photometric quality.
Furthermore, they cover the orbit better, especially with respect to the 
shoulders of the minima, and this allows more accurate photo\-me\-tric 
elements to be determined. 
The eclipses have been covered several times and most out--of--eclipse 
phases at least twice.
In addition, eclipse observations were made January - February 2002,
October 2005, and December 2006 in order to establish more times of minima 
and investigate whether apsidal motion could be detected; see below. 

\subsection{Ephemeris and apsidal motion}
\label{sec:swcma_eph}

Five times of primary minimum and six of secondary minimum, derived
from the $uvby$ observations, are given in Table~\ref{tab:swcma_tmin}
together with 26 published times.
Weighted (weight proportional to the square of the inverse rms) least squares fits to
all times of primary and secondary mi\-ni\-ma yield nearly identical periods of
$10\fd 091985 \pm 0\fd 000002$ and $10\fd 091981 \pm 0\fd 000003$, respectively,
whereas adopting equal weights leads to slightly larger values of 
$10\fd 092003 \pm 0\fd 000008$ and $10\fd 091999 \pm 0\fd 000005$.
However, the derived linear ephemerides do not represent the observed times 
very well.
Ziegler (\cite{ziegler65}) derived $10 \fd 091948$, which was adopted
by Lacy (\cite{lacy97b}).
If only the photoelectric times of minima are fitted, 
periods of $10\fd 091983 \pm 0\fd 000001$ and $10\fd 091977 \pm 0\fd 000002$
are obtained from primary and se\-con\-dary eclipses, respectively. 

\begin{figure}
\epsfxsize=85mm
\epsfbox{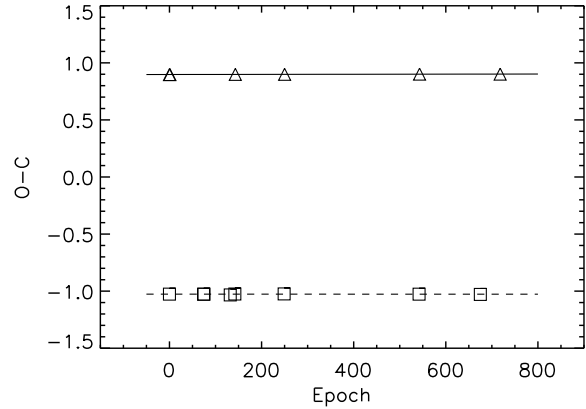}
\caption[]{\label{fig:swcma_aps_pe}
Apsidal motion for SW\,CMa. 
O-C are the residuals (days)
from the linear part of the apsidal motion ephemeris defined by the
parameters given in Table \ref{tab:swcma_aps}.
The full and dashed curves represent predictions
from the apsidal motion parameters for primary (triangles)
and secondary (squares) eclipses, respectively. Only the
photoelectric observations are shown.
}
\end{figure}

Since the orbit of SW\,CMA is eccentric ($e \approx 0.3$), 
apsidal motion should be present, as also discussed by Lacy 
(\cite{lacy97b}), who found a poorly determined apsidal motion period 
of several thousand years.
We now have more precise times of minima available and 
derive the apsidal motion parameters presented in Table~\ref{tab:swcma_aps}
from a weighted least squares method, following the formalism by 
Gim\'enez \& Garcia-Pelayo (\cite{ggp83}) and Gim\'enez \& Bastero 
(\cite{gb95}). The orbital inclination $i$ and
eccentricity $e$ were fixed to the values derived from the photometric
analysis (Torres et al., in prep.). As seen, 
also in Fig.~\ref{fig:swcma_aps_pe}, a slow but sig\-ni\-fi\-cant motion
has been detected, but the apsidal motion period is still very uncertain,
although the value obtained is in good agreement with predictions from
stellar models.
For more precise apsidal motion information, SW\,CMa should be monitored 
regularly in the coming decades.

The detected apsidal motion leads to an insignificant phase shift of the secondary
eclipse compared to the primary eclipse of only $-0.00012$ during the 1987--92 
$uvby$ light curve observations. Therefore, a linear ephemeris can
safely be used for the light curve analysis, and we adopt 

\begin{equation}
\label{eq:swcma_eph}
\begin{tabular}{r r c r r}
{\rm Min \, I} =  & 2446829.6482 & + & $10\fd 091988$ &$\times\; E$ \\
                  &      $\pm  1$&   &       $\pm  5$ &             \\
\end{tabular}
\end{equation}

\noindent
which has been derived from the 1982--87 times of mi\-ni\-ma.
Secondary eclipses occur at phase 0.309. 
For com\-pa\-ri\-son, the method of Lafler \& Kinman (\cite{lk65}) applied 
to the 1987--92 $uvby$ eclipse observations gives a period of $10 \fd 091982$. 

The 2002--2006 $uvby$ eclipse observations have not been included in the 
light curve analyses, since their phase shifts are larger. 

\begin{table}
\caption[]{\label{tab:swcma_aps}
Apsidal motion parameters for SW\,CMa derived from the
photoelectric times of minima. 
}
\begin{center}
\begin{tabular}{ll} \hline
\hline\noalign{\smallskip}
Parameter    & Value and rms error                    \\
\noalign{\smallskip}
\hline
\noalign{\smallskip}
$i$ (\degr)                & 88.6 (assumed)             \\           
$e$                        & 0.316 (assumed)            \\
$T_0$                      & $2446828.7512 \pm 0.0006 $  \\
$P_{anomalistic}$ (d)      & $10.091997 \pm 0.000005$     \\ 
$P_{sidereal}$    (d)      &  10.091978                  \\ 
$\omega_0$ (\degr)         & $163.52 \pm 0.10$           \\
$\omega_1$ (\degr / cycle) & $0.00067 \pm 0.00021$       \\
$U$ (yr)                   & $14900 \pm 3600 $           \\
\hline
\end{tabular}
\end{center}
\end{table}

\section{\object{V636\,Cen}} 
\label{sec:v636cen}

\begin{figure*}
\epsfxsize=185mm
\epsfbox{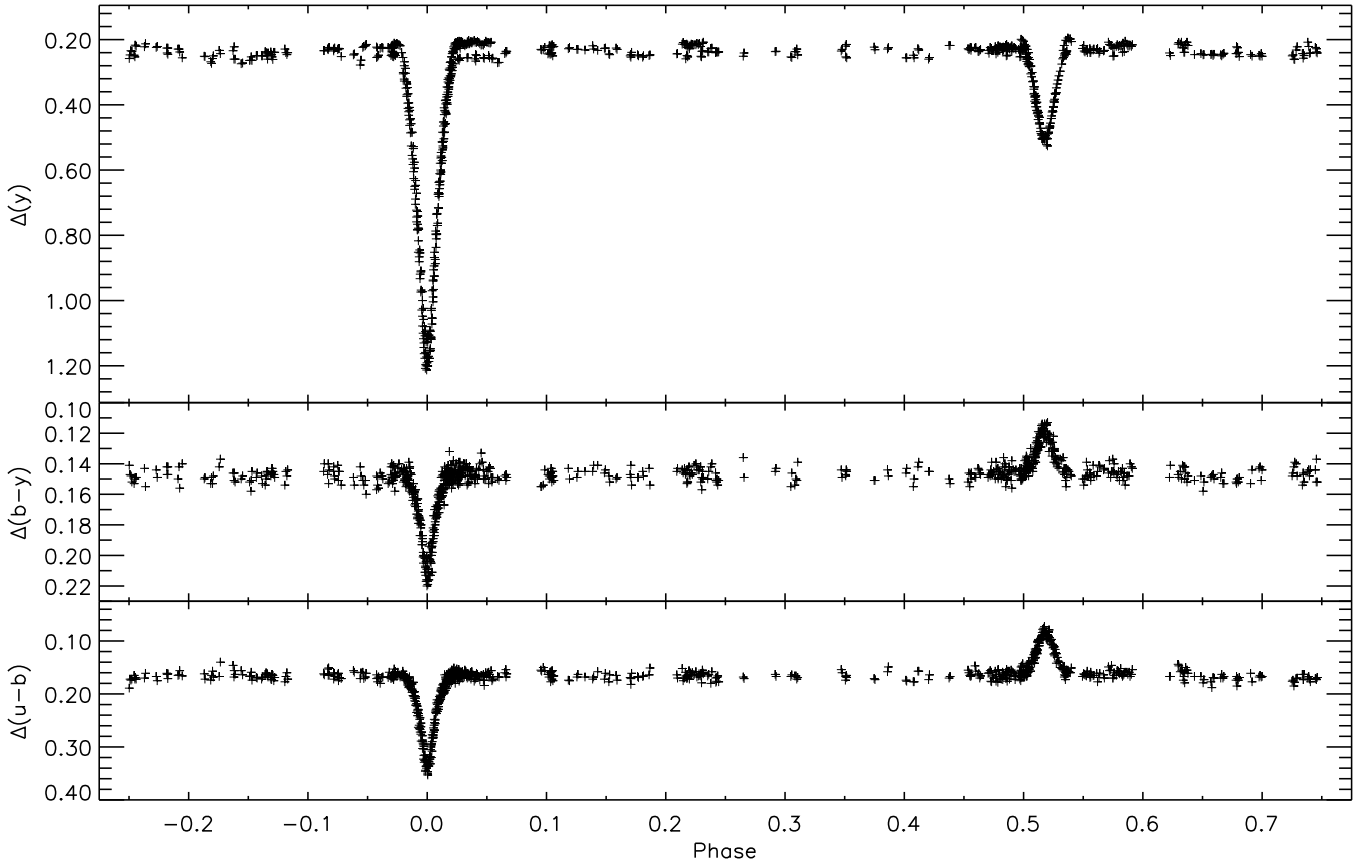}
\caption[]{\label{fig:v636cen}
$y$ light curve and $b-y$ and $u-b$ colour curves (instrumental system)
for V636\,Cen.
}
\end{figure*}

From 276 visual (1952--53) and 282 photographic (1935--38, 1952--53) 
observations, Hoffmeister (\cite{h58}) reported V636\,Cen = HD124784 = HIP69781
to be a $4 \fd 3$ period G0 eclipsing binary, and he gave a linear ephemeris 
and seven times of minima. Later, Popper (\cite{dmp66}) obtained two 22 
{\AA}/mm spectra and found sharp but single lines. 
Today V636\,Cen is known to be double-lined, but the lines of the 
secondary component are much fainter than those from the primary. 
Dworak (\cite{d73}) pointed out that V636\,Cen is within 100 pc of the Sun, 
and it was included in the list by Dworak \& Oblak (\cite{do89}) of 95 
such eclipsing binaries selected for Hipparcos observations.
The parallax determined by Hipparcos (ESA \cite{hip97}) is 
$15.36 \pm 1.12$ mas, corresponding to a distance of only 65 pc. 
With accurate absolute dimensions becoming available, V636\,Cen can 
therefore be added to the (short) list of eclipsing binaries useful for 
the definition of the radiative flux scale (Popper \cite{dmp98b}).

A preliminary analysis of V636\,Cen, based on the $uvby$ light curves 
presented here and complete spectroscopic orbits from CORAVEL radial velocity 
measurements has been done by Larsen (\cite{al98}). She finds that V636\,Cen 
is one of the few well studied systems with components in the 
$0.7$--$1.1$ $M_{\sun}$ range;
preliminary masses are 1.04 and 0.85 $M_{\sun}$, respectively.
Furthermore, V636\,Cen is well detached, and such systems are important
for detailed tests of stellar models around and especially below 1 $M_{\sun}$.

\subsection{$uvby$ light curves}
\label{sec:v636cen_lc}

$uvby$ light curves, containing in total 853 points in each colour, were
observed on 76 nights during six periods between March 1985 and April 1991
(JD2446144--JD2448377).
\object{HD124829} and \object{HD125444} were selected as comparison stars,
see Tables~\ref{tab:comparisons_dm}, \ref{tab:comparisons_inf}, and
\ref{tab:std_publ} for further information, and they were both found to be 
constant within the observational accuracy of a few mmag during our 
observations.

The light curves of V636\,Cen are shown in Fig.~\ref{fig:v636cen}, and 
variations due to surface activity (from spots) are clearly seen from the 
$uvby$ observations. However, the large number of observations obtained 
during some of the seasons allows a detailed study of the activity and its 
influence on the accuracy of the photometric elements (Larsen \cite{al98}). 
As seen, V636\,Cen consists of two well detached components of very different 
surface fluxes in an eccentric orbit ($e \approx 0.13$).

Further eclipse observation were made January 2002--July 2007. They
have been used in the apsidal motion ana\-ly\-sis presented below, but not in 
the light curve ana\-ly\-ses.

\subsection{Ephemeris and apsidal motion}
\label{sec:v636cen__eph}

Nine times of primary and nine of secondary minimum, derived
from the $uvby$ observations, are given in Table~\ref{tab:v636cen_tmin}
together with eight published times of primary minimum (none available for
the secondary minimum). 
Weighted (weight proportional to square of inverse rms) least squares 
fits to all times of primary and secondary minima yield significantly
different periods of
$4\fd 28394311 \pm 0\fd 00000013$ and
$4\fd 28394845 \pm 0\fd 00000019$, respectively.
Fitting only the $uvby$ times of primary minima leads to the same
period.

As mentioned in Section~\ref{sec:v636cen_lc}, the orbit of V636\,Cen is
eccentric ($e = 0.13$), and the secondary eclipse occurs near phase 0.52. 
Apsidal motion must therefore
exist, although at a slow rate, since the relative radii of the
components are small. From theoretical density concentration
coefficients, the absolute dimensions, the eccentricity
and inclination of the orbit, and assuming the rotation of the components
to be pseudosynchronized, Larsen (\cite{al98}) derived an expected apsidal 
motion period of $U \approx 6\,000$ yr. The relativistic contribution and the
classical contribution from tidal effects are of almost the same
size ($P/U \approx 10^{-6}$), making V636\,Cen particularly interesting among
apsidal motion systems.

\begin{figure}
\epsfxsize=85mm
\epsfbox{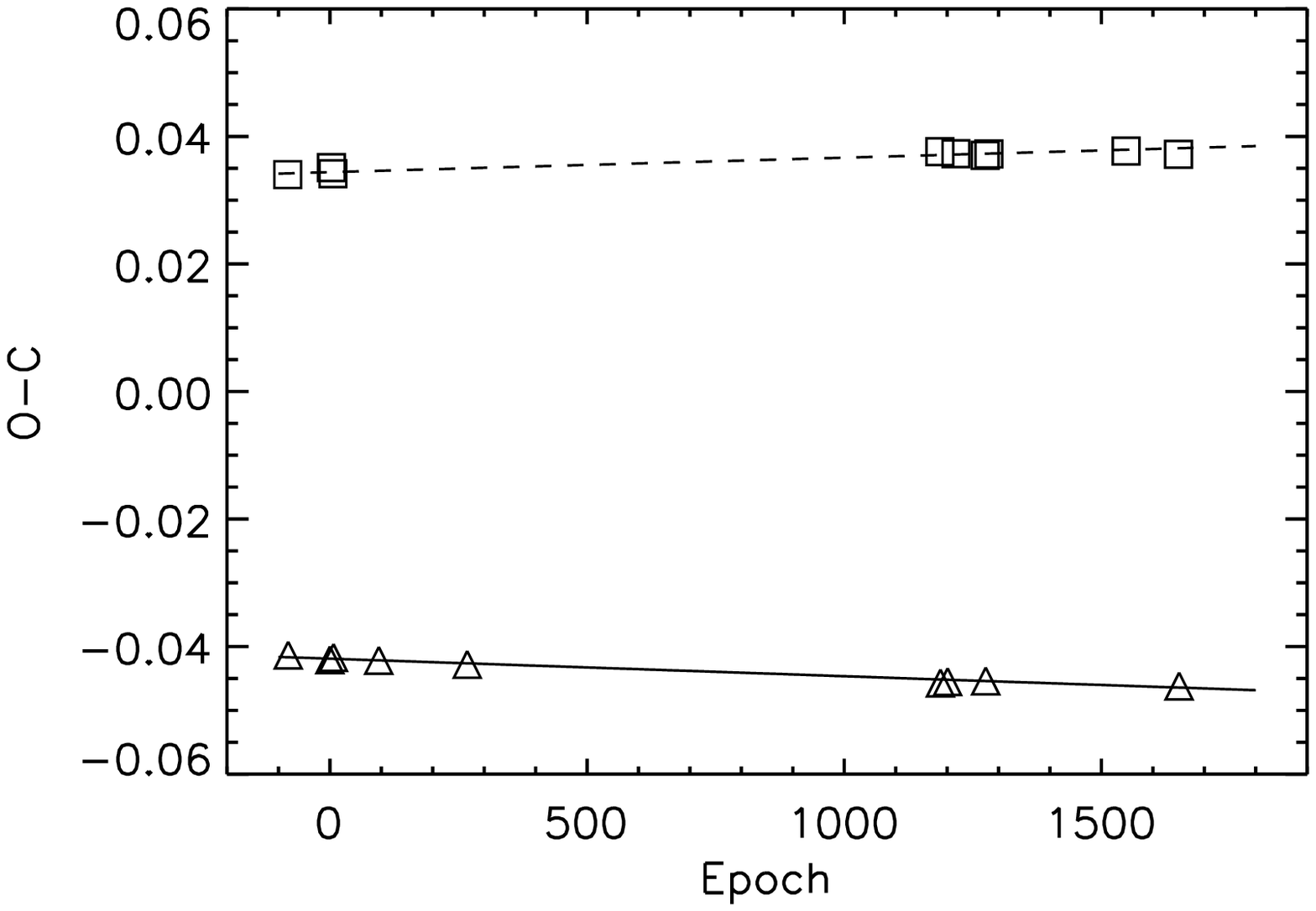}
\caption[]{\label{fig:v636cen_aps}
Apsidal motion for V636\,Cen.
O-C are the residuals (days)
from the linear part of the apsidal motion ephemeris defined by the
parameters given in Table \ref{tab:v636cen_aps}.
The full and dashed curves represent predictions
from the apsidal motion parameters for primary (triangles)
and secondary (squares) eclipses, respectively. Only the
$uvby$ observations are shown.
}
\end{figure}

With additional $uvby$ times of minima available, we have
repeated the apsidal motion analysis and obtain the parameters listed
in Table~\ref{tab:v636cen_aps}; see also Fig.~\ref{fig:v636cen_aps}. 
Well defined apsidal motion parameters have now been determined for
meaningful comparison with theoretical density concentrations of
stellar models.

The apsidal motion does not lead to significant phase shifts of secondary
eclipse compared to primary eclipse during the 1985--1991 $uvby$ light
curve observations. We have adopted the following linear ephemeris,
derived from the corresponding times of primary minima for the light
curve analyses:

\begin{equation}
\label{eq:v636cen_eph}
\begin{tabular}{r r c r r}
{\rm Min \, I} =  & 2446873.80124 & + & $4\fd 2839423$ &$\times\; E$
\\
                  &      $\pm  14$&   &     $\pm    7$ &   \\
\end{tabular}
\end{equation}

\begin{table}
\caption[]{\label{tab:v636cen_aps}
Apsidal motion parameters for V636\,Cen derived from the
photoelectric times of minima. 
}
\begin{center}
\begin{tabular}{ll} \hline
\hline\noalign{\smallskip}
Parameter    & Value and rms error                    \\
\noalign{\smallskip}
\hline
\noalign{\smallskip}
$i$ (\degr)                & 89.65 (assumed)             \\           
$e$                        & 0.134 (assumed)            \\
$T_0$                      & $2447220.8424 \pm 0.0002 $  \\
$P_{anomalistic}$ (d)      & $4.283956 \pm 0.0000008$    \\ 
$P_{sidereal}$    (d)      &  4.283946                   \\ 
$\omega_0$ (\degr)         & $281.95 \pm 0.06$           \\
$\omega_1$ (\degr / cycle) & $0.00080 \pm 0.00005$       \\
$U$ (yr)                   & $5270 \pm 335 $           \\
\hline
\end{tabular}
\end{center}
\end{table}

\section{\object{VZ\,Hya}} 
\label{sec:vzhya}

\begin{figure*}
\epsfxsize=185mm
\epsfbox{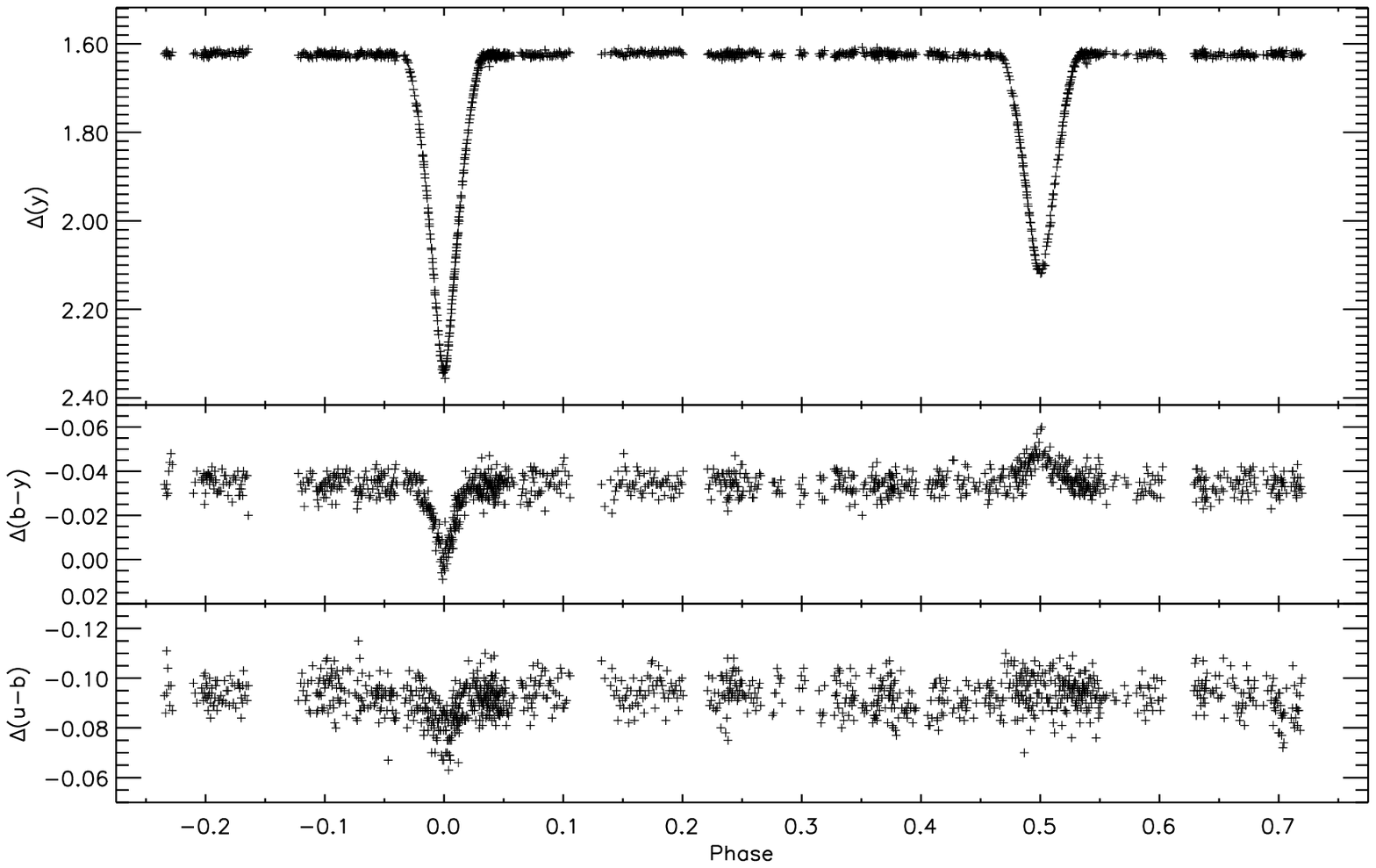}
\caption[]{\label{fig:vzhya}
$y$ light curve and $b-y$ and $u-b$ colour curves (instrumental system)
for VZ\,Hya.
}
\end{figure*}

VZ\,Hya =  HD72257 = HIP41834 was discovered by O'Connell (\cite{oconnell32}) 
to be an eclipsing binary, and spectroscopic elements were obtained from 
75 {\AA}/mm spectra by Struve (\cite{struve45}), who showed that the period is 
twice the value given by O'Connell.
Early light curves and absolute dimensions were published by Wood 
(\cite{wood46}, visual data) as part of his thesis, and by Gaposchkin 
(\cite{gap53}, photographic data). 
From  20 and 40 {\AA}/mm spectra, Popper (\cite{dmp65}) later determined 
improved elements, leading to masses accurate to about 2.5\%. 

The first photoelectric ($UBV$) light curves were presented by Walker 
(\cite{walker70}) together with a preliminary photometric solution, whereas 
Wood (\cite{wood71}) performed a complete analysis as part of the introduction of his
then new binary model and the corresponding WINK code. 
A photometric analysis, also based on WINK, was published by Cester et al. 
(\cite{cester78}).

Another set of photoelectric $UBV$ light curves were published by 
Padalia \& Srivastava (\cite{ps75}). In contrast to Walker, Wood, and Cester 
et al. they found primary eclipse to be an occultation (larger star in front) 
rather than a transit (smaller star in front). 
This picture of VZ\,Hya was soon shown by Popper (\cite{dmp76}) to be wrong, 
but it was nevertheless defended once more by Padalia (\cite{p86}). 

Because its dimensions were relatively well known, VZ\,Hya has often
been included in binary samples used in various astrophysical 
investigations, e.g. He abundance in Population I stars 
(Popper et al. \cite{dmpetal70}), stellar radii (Lacy \cite{lacy77}, 
Shallis \& Blackwell \cite{sb80}, Pastori et al. \cite{petal85}), 
and mass-radius relations (Zhai \& Zhang \cite{zz89}).
The currently most reliable absolute dimensions are those given by 
Popper (\cite{dmp80}) in his critical review of stellar masses. There is, 
however, still ample space for improvement, provided more accurate light 
curves and radial velocities are obtained. Since VZ\,Hya is an important 
F-type system, we decided to obtain the observations needed for this
purpose.

\subsection{$uvby$ light curves}
\label{sec:vzhya_lc}
 
Complete $uvby$ light curves containing 1180 points in each colour were
observed on 44 nights during four periods between February 1989 and April 1992 
(JD2447575--JD2448684).
\object{HD72528}, \object{HD71615}, and \object{HD72782} were used as
comparison stars and were found to be 
constant within the observational accuracy of a few mmag during our 
observations; see Tables~\ref{tab:comparisons_dm}, \ref{tab:comparisons_inf}, 
and \ref{tab:std_publ} for further information. 

The light curves of \VZ\ are shown in Fig.~\ref{fig:vzhya}.
As seen, VZ\,Hya consists of two well-detached components of different surface
fluxes in a circular orbit.
Both eclipses have been covered several times and most out-of-eclipse phases
at least twice. The accuracy per point is about 0.004 mag ($vby$) and 
0.006 mag ($u$), meaning that the new light curves are of significantly higher 
quality than those by Walker (\cite{walker70}) and Padalia \& Srivastava 
(\cite{ps75}). 
We see no signs of intrinsic variability.
Photometric analyses reveal that the secondary eclipse is very close to being total
(Clausen et al., \cite{avw08}).

\begin{figure*}
\epsfxsize=185mm
\epsfbox{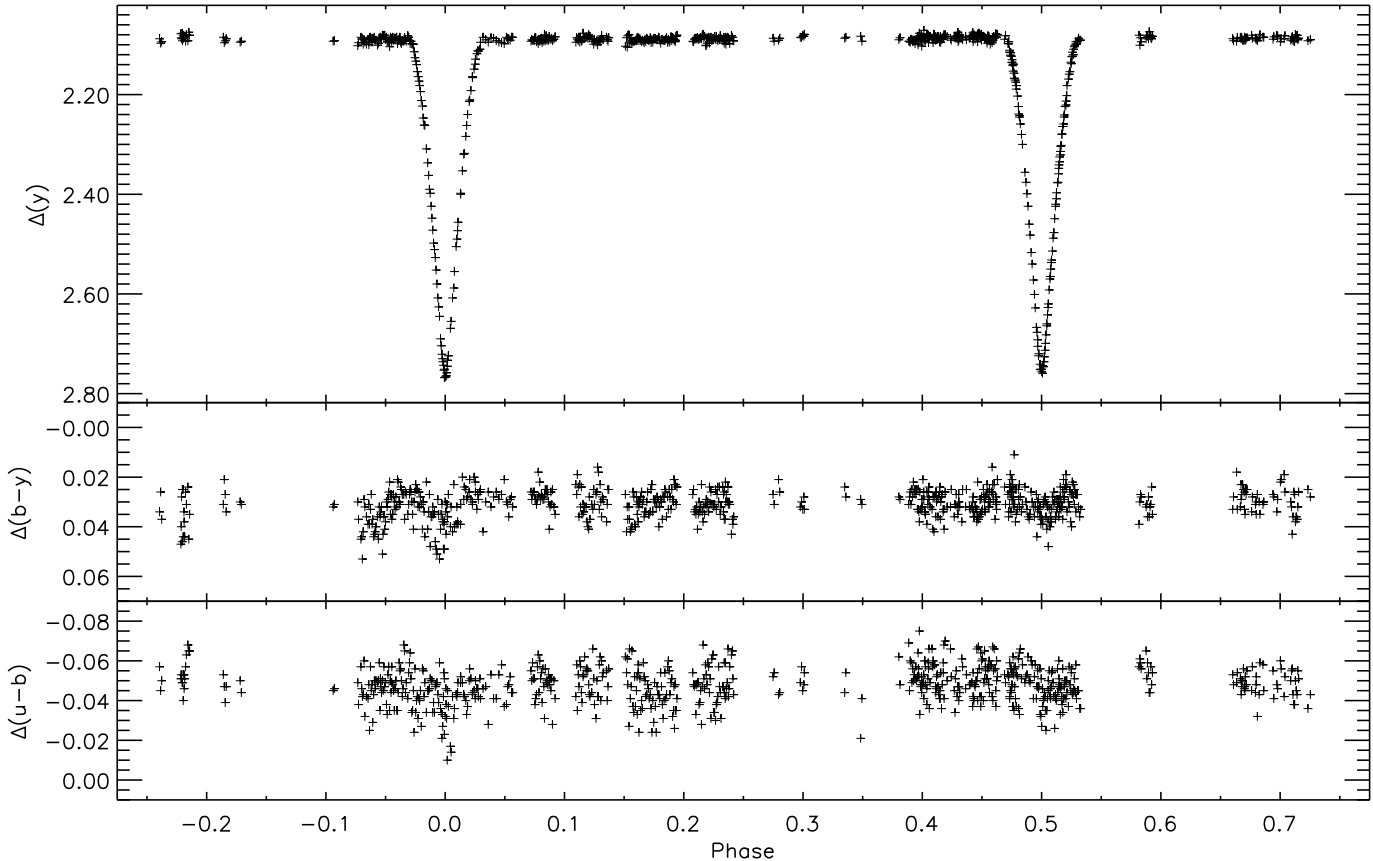}
\caption[]{\label{fig:wzoph}
$y$ light curve and $b-y$ and $u-b$ colour curves (instrumental system)
for WZ\,Oph.
}
\end{figure*}

\subsection{Ephemeris}
\label{sec:vzhya_eph}

Three times of primary minimum and one of secondary minimum, derived
from the $uvby$ observations, are given in Table~\ref{tab:vzhya_tmin}
together with 26 published times and times redetermined from the 
photometry by Walker (\cite{walker70}).

We adopt the following linear ephemeris, which is based on all available times
assuming a circular orbit and derived as described in Sect.~\ref{sec:adboo_eph}:

\begin{equation}
\label{eq:vzhya_eph}
\begin{tabular}{r r c r r}
{\rm Min \, I} =  & 2448273.63450 & + & $2\fd 90430023$ &$\times\; E$ \\
                  &      $\pm   9$&   &       $\pm  11$ &             \\
\end{tabular}
\end{equation}

\noindent
Within the errors, separate least squares fits to all primary and secondary 
times yield identical periods.
The period is close to the most recent published determination  
(Srivastava \cite{s87}); 
see also Kreiner\footnote{http://www.as.ap.krakow.pl/ephem}.
Srivastava found evidence for a slowly increasing period and sinusoidal 
variations of the $O-C$ residuals, in\-di\-ca\-ting a possible third body 
in the system. 
The data in Table \ref{tab:vzhya_tmin} provide no clear evidence for such
variations, however.

\section{\object{WZ\,Oph}} 
\label{sec:wzoph}

WZ\,Oph = HD154676 = HIP83719 was discovered by J.H. Metcalf to be an 
eclipsing binary (Pickering \cite{p17}). Visual light curves and times of 
minima were published by Leiner (\cite{leiner21}, \cite{leiner26}), who
derived an orbital period of about $2 \fd 09$, whereas the correct value 
of about $4 \fd 18$ was proposed by McLaughlin (\cite{mcl29}). 
Lause (\cite{lause36}) presented further visual minima.
Sanford (\cite{s37}) obtained the first radial velocities, and Gaposchkin
(\cite{gap38}) established visual (from Leiner's observations) and photographic 
(from Harvard patrol plates) light curves, and he derived the first absolute 
dimensions of the two nearly identical components of WZ\,Oph.

The first modern analysis of WZ\,Oph, leading to masses and radii accurate to
$3$--$4$\%, was done by Popper (\cite{dmp65}) from photoelectric light curves 
and 20 {\AA}/mm spectra. WZ\,Oph is included in his comprehensive review on 
stellar masses (Popper \cite{dmp80}), but not in the later review on accurate 
masses of radii of normal stars by Andersen (\cite{ja91}), where a more strict 
limit of 2\% was applied. Very accurate light curves and radial velocities are 
needed in order to reach such accuracies (e.g. Andersen et al. 
\cite{jaetal80}), and besides standard $uvby$ indices (Hilditch \& Hill 
\cite{hh75}) and additional times of minima (see Tables~\ref{tab:wzoph_tminp}, 
\ref{tab:wzoph_tmins}), practically no new observational data for WZ\,Oph have
been published since those by Popper. We now remedy this lack.

\subsection{$uvby$ light curves}
\label{sec:wzoph_lc}

Complete $uvby$ light curves containing 697 points in each colour were
observed on 41 nights during six periods between May 1991 and March 1997
(JD2448370--JD2450535).
\object{HD155193} and \object{HD154931} were selected as comparison stars,
and they were found to be constant within 
the observational accuracy of a few mmag during our observations; 
see Tables~\ref{tab:comparisons_dm}, \ref{tab:comparisons_inf}, and
\ref{tab:std_publ}  for further information.
The eclipses have been covered on several nights and out-of-eclipse 
phases, except for a few gaps, at least twice.

As seen in Fig.~\ref{fig:wzoph}, WZ\,Oph consists of two well-detached 
components of almost identical surface fluxes in a circular orbit, and from 
the analysis by Popper (\cite{dmp65}) they are also known to be very similar 
with respect to masses and radii. 
The accuracy per point is about 0.004 mag ($vby$) and 0.006 mag ($u$), 
meaning that they are of significantly higher quality than those by 
Popper (\cite{dmp65}).
However, our photometric analyses (Clausen et al., \cite{avw08}) reveal 
that throughout all phases the points scatter by 1--4 mmag more than this, 
highest in $u$. At a given phase, the observations from different seasons 
do not differ sy\-ste\-ma\-tical\-ly.
The extra scatter may be due to low-level surface activity, since
weak emission features in the Ca~II H and K lines are seen at the position of 
both components on high-resolution spectra of WZ\,Oph.

For the ephemeris adopted by Popper, the component eclipsed at
phase 0.0 (his component 1) was found to be (formally) slightly more massive,
slightly smaller, and appeared to have slightly weaker lines and to be less 
luminous than the other. 
The new accurate $uvby$ light curves reveal that, adopting Popper's ephemeris,
the eclipse at phase 0.5 is in fact slightly deeper in all four colours, 
meaning that the component eclipsed here has a slightly
higher surface flux than the other. Since analyses of 
new CfA radial velocities (Clausen et al., \cite{avw08}) show
that this component is also marginally the more massive, we have 
decided to interchange the minima compared to Popper's notation, 
so for the linear ephemeris presented below, the deeper 
(primary) eclipse occurs at phase 0.0.

\subsection{Ephemeris}
\label{sec:wzoph_eph}

From the $uvby$ observations, two times of primary and two times of
secondary minimum are available, see Tables~\ref{tab:wzoph_tminp}
and \ref{tab:wzoph_tmins}, where the 104 times of minima
we have been able to locate in the literature are also included.
We note that those by Gaposchkin (\cite{gap38}) are based on Harvard patrol 
plates with usual exposure times of one hour.

We adopt the following linear ephemeris, which is based on all available times,
assuming a circular orbit, and derived as described in 
Sect.~\ref{sec:adboo_eph}:

\begin{equation}
\label{eq:wzoph_eph}
\begin{tabular}{r r c r r}
{\rm Min \, I} =  & 2450535.78331 & + & $4\fd 18350681$ &$\times\; E$ \\
                  &      $\pm  24$&   &       $\pm  24$ &             \\
\end{tabular}
\end{equation}
\noindent

\noindent
It represents a significant improvement compared to previous determinations.
Excluding the rather inaccurate times by Gaposchkin (\cite{gap38}), 
which also show large $O-C$ residuals in Tables~\ref{tab:wzoph_tminp} and
\ref{tab:wzoph_tmins}, does not change the result significantly.  
The Lafler \& Kinman (\cite{lk65}) method applied to
the $uvby$ observations yields $4 \fd 1835070$. 
Finally, separate least squares fits to all primary and secondary times 
yield identical periods within errors.

Note that, as mentioned above, the new epoch is shifted by half a period 
compared to (most) previous ephemerides given for WZ\,Oph, including that by 
Kreiner\footnote{http://www.as.ap.krakow.pl/ephem}.

\section{Summary and conclusions}
\label{sec:conclusions}

Complete $uvby$ light cuves and $uvby\beta$ standard indices have been
obtained for the six main-sequence eclipsing binaries \AD, \HW, \SW, \V636, \VZ,
and \WZ. For \V636\ and \HW, the light curves are the first modern ones obtained,
and for the other systems they are both more complete and more accurate than
earlier data. Ephemerides have been determined from all available times of
minima, and apsidal motion has been detected and quantified for \SW\ and
\V636.

For \AD, the new photometry indicates low-level surface activity,
and spectroscopy supports that the cooler secondary component is the most
likely candidate (Clausen et al. \cite{avw08}).
The new photometry and spectroscopy also suggests that \WZ\ is
mildly active. 
The $uvby$ light curves of \V636\ clearly show variations due to
activity at a moderate level, also confirmed by spectroscopy (Clausen et al.
in prep.). 

In parallel, new spectroscopic observations for accurate radial velocities
and abundance determinations were obtained. The combined data yield
accurate absolute dimensions and heavy-element abundances for all six
systems and allow detailed tests of current stellar evolu\-tio\-na\-ry 
mo\-dels. Our analyses will be published in three separate papers 
(Clausen et al. \cite{avw08}; the F-type systems \AD, \VZ, and \WZ),
(Clausen et al. in prep.; the G-type system \V636);
(Torres et al. in prep.; the late A-type systems \HW\ and \SW).

\begin{acknowledgements}

R. Clement, O. Liu, and M. Viskum kindly observed some of
the binaries on a few nights as part of their own programs at SAT.
Excellent technical support was received from the staff of Copenhagen 
University and from the staff at ESO, La Silla.
We thank J.M. Kreiner for providing complete lists of times of eclipse for
the binaries, and G. Torres for refining the ephemeris calculations for
\AD, \VZ, and \WZ.
The CORAVEL and CfA groups kindly made unpublished 
radial velocity information and preliminary spectroscopic orbits available
for the preparation of this paper.

The projects "Structure and evolution of stars - new insight from
eclipsing binaries and pulsating stars" and "Stars: Central engines of 
the evolution of the Universe", carried out at Copenhagen University
and Aarhus University, are supported by the Danish National Science
Research Council.  
LPRV greatfully acknowledges partial support from the Brazilian
agencies CNPq, CAPES and FAPEMIG.
Furthermore, this investigation was supported by the Danish Board 
for Astronomical Research, the Carlsberg Foundation (to JA), and the 
Spanish Science Research Council.

The following internet-based resources were used in research for
this paper: the NASA Astrophysics Data System; the SIMBAD database
and the VizieR service operated by CDS, Strasbourg, France; the
ar$\chi$iv scientific paper preprint service operated by Cornell University;
the BBSAG and Orion bulletins (http://www.bav-astro.de);
the VSOLJ Variable Star Bulletin (http://vsolj.cetus-net.org);
the AASVO archive (http://www.aavso.org).
\end{acknowledgements}

{}

\listofobjects

\begin{appendix}
\section{Times of minima tables}

\onecolumn
\begin{longtable}{lllrlc}
\caption[]{\label{tab:adboo_tminp}
Times of $primary$ minima for AD\,Boo.
O-C values are calculated for the
ephemeris given in Eq.  \ref{eq:adboo_eph} assuming a circular orbit. 
References are:
B = BBSAG Bulletin, followed by volume number and year of publication.
BA78 = Baldwin (\cite{ba78}).
BS97 = Baldwin \& Samolyk~(\cite{bs97}).
D05 = Diethelm (\cite{d05}).
H06 = H\"ubscher et al. (\cite{h06}).
I86 = Isles (\cite{i86}).
MK04 = Maciejewski \& Karska (\cite{mk04}).
N04 = Nagai (\cite{n04}).
N05 = Nagai (\cite{n05}).
O = Orion (Diethelm \& Locker \cite{dl70a}, \cite{dl71a}, \cite{dl71b}, \cite{dl71c}), 
followed by volume number and year of publication.
S63 = Strohmeier et al. (\cite{sm63}).
S68 = Strohmeier et al. (\cite{sm68}).
VB74 = van Buren (\cite{vb74}); his data are taken from B and O.
Z83 = Zhai et al. (\cite{zhai83}).
Observing methods are: PG photographic; V visual; PE photoelectric or CCD.
Times marked with an asterisk were published after we completed the
ephemeris calculations; including them makes no significant changes.
}\\
\hline\hline\noalign{\smallskip}
HJD          & rms     &  rms     &   O-C     &Observing&  Reference  \\
$-$ 2\,400\,000& publ. &  adopt.  & days      &method   &             \\
\noalign{\smallskip}
\endfirsthead
\caption{continued.}\\
\hline\hline\noalign{\smallskip}
HJD          & rms     &  rms     &   O-C     &Observing&  Reference  \\
$-$ 2\,400\,000& publ.  &  adopt.  & days     &method   &             \\
\hline\noalign{\smallskip}
\endhead
\hline
\noalign{\smallskip}
  15552.681  &         &   0.0320 &    0.0426 &   PG &  S68        \\ 
  16576.720  &         &   0.0320 &    0.0221 &   PG &  -          \\ 
  16605.709  &         &   0.0320 &    0.0478 &   PG &  -          \\ 
  16880.830  &         &   0.0320 &    0.0175 &   PG &  -          \\ 
  17329.796  &         &   0.0320 &    0.0524 &   PG &  -          \\ 
  17660.792  &         &   0.0320 &    0.0393 &   PG &  -          \\ 
  19439.923  &         &   0.0320 & $-$0.0038 &   PG &  -          \\ 
  20244.712  &         &   0.0320 &    0.0193 &   PG &  -          \\ 
  23612.695  &         &   0.0320 & $-$0.0156 &   PG &  -          \\ 
  23612.706  &         &   0.0320 & $-$0.0046 &   PG &  -          \\ 
  26190.436  &         &   0.0320 & $-$0.0082 &   PG &  S63        \\ 
  26767.648  &         &   0.0320 &    0.0067 &   PG &  -          \\ 
  27183.483  &         &   0.0320 &    0.0114 &   PG &  -          \\ 
  27212.436  &         &   0.0320 &    0.0011 &   PG &  -          \\ 
  27609.605  &         &   0.0320 & $-$0.0408 &   PG &  S68        \\ 
  28219.931  &         &   0.0320 & $-$0.0129 &   PG &  -          \\ 
  28962.610  &         &   0.0320 & $-$0.0356 &   PG &  S63        \\ 
  29014.430  &         &   0.0320 &    0.0642 &   PG &  -          \\ 
  30439.803  &         &   0.0320 &    0.0292 &   PG &  S68        \\ 
  32659.663  &         &   0.0320 &    0.0592 &   PG &  -          \\ 
  33023.667  &         &   0.0320 & $-$0.0468 &   PG &  -          \\ 
  37016.535  &         &   0.0320 &    0.0236 &   PG &  S63        \\ 
  37018.539  &         &   0.0320 & $-$0.0412 &   PG &  -          \\ 
  37018.585  &         &   0.0320 &    0.0048 &   PG &  -          \\ 
  37351.647  &         &   0.0320 & $-$0.0112 &   PG &  -          \\ 
  40711.396  &         &   0.0150 & $-$0.0048 &   V  &  O119(1970),VB74 \\ 
  41042.409  &         &   0.0150 & $-$0.0009 &   V  &  O124(1971) \\ 
  41042.411  &         &   0.0150 &    0.0011 &   V  &  -          \\ 
  41048.614  &         &   0.0150 & $-$0.0024 &   V  &  -          \\ 
  41071.381  &         &   0.0150 &    0.0078 &   V  &  O125(1971),VB74 \\ 
  41104.453  &         &   0.0150 & $-$0.0211 &   V  &  -          \\ 
  41135.498  &         &   0.0150 & $-$0.0083 &   V  &  O126(1971) \\ 
  41135.503  &         &   0.0150 & $-$0.0033 &   V  &  -          \\ 
  41162.412  &         &   0.0150 &    0.0113 &   V  &  -          \\ 
  41402.385  &         &   0.0150 &    0.0026 &   V  &  B3(1972),VB74 \\ 
  41402.393  &         &   0.0150 &    0.0106 &   V  &  B4(1972),VB74 \\ 
  41433.423  &         &   0.0150 &    0.0085 &   V  &  B3(1972),VB74 \\ 
  41764.378  &         &   0.0150 & $-$0.0456 &   V  &  B8(1973),VB74 \\ 
  41764.413  &         &   0.0150 & $-$0.0106 &   V  &  -          \\ 
  41795.459  &         &   0.0150 &    0.0033 &   V  &  B9(1973),VB74 \\ 
  41824.387  &         &   0.0150 & $-$0.0320 &   V  &  -          \\ 
  41853.381  &         &   0.0150 & $-$0.0013 &   V  &  B10(1973),VB74 \\ 
  42186.454  &         &   0.0150 & $-$0.0062 &   V  &  B15(1974)  \\ 
  42275.417  &         &   0.0150 & $-$0.0019 &   V  &  B17(1974)  \\ 
  42614.686  &         &   0.0150 & $-$0.0173 &   V  &  BA78       \\ 
  42860.8760 &         &   0.0150 & $-$0.0153 &   V  &  BS97       \\ 
  42937.427  &         &   0.0150 & $-$0.0102 &   V  &  B28(1976)  \\ 
  42997.442  &         &   0.0150 &    0.0094 &   V  &  B29(1976)  \\ 
  44079.417  &         &   0.0150 & $-$0.0017 &   V  &  B44(1979)  \\ 
  44704.1985 & 0.0005  &   0.0007 &    0.0001 &   PE &  Z83        \\ 
  44731.0935 & 0.0005  &   0.0007 &    0.0006 &   PE &  -          \\ 
  44766.2627 & 0.0003  &   0.0004 &    0.0001 &   PE &  -          \\ 
  45074.5133 & 0.0005  &   0.0035 & $-$0.0016 &   PE &  B60(1982)  \\ 
  45082.781  &         &   0.0150 & $-$0.0091 &   V  &  BS97       \\ 
  45101.407  &         &   0.0150 & $-$0.0024 &   V  &  B60(1982)  \\ 
  45442.761  &         &   0.0150 & $-$0.0015 &   V  &  BS97       \\ 
  46216.495  &         &   0.0150 & $-$0.0014 &   V  &  I86        \\ 
  46216.508  &         &   0.0150 &    0.0117 &   V  &  -          \\ 
  46907.472  &         &   0.0150 & $-$0.0059 &   V  &  -          \\ 
  46963.339  &         &   0.0150 &    0.0033 &   V  &  B84(1987)  \\ 
  47161.910  &         &   0.0150 & $-$0.0312 &   V  &  BS97       \\ 
  47219.865  &         &   0.0150 & $-$0.0028 &   V  &  -          \\ 
  47246.7620 & 0.0010  &   0.0014 & $-$0.0003 &   PE &  This paper \\ 
  47248.8312 & 0.0002  &   0.0003 &    0.0001 &   PE &  -          \\ 
  47383.313  &         &   0.0150 &    0.0095 &   V  &  B89(1988)  \\ 
  47579.8410 & 0.0010  &   0.0014 &    0.0008 &   PE &  This paper \\ 
  48026.7022 & 0.0002  &   0.0003 & $-$0.0003 &   PE &  -          \\ 
  48357.704  &         &   0.0150 & $-$0.0076 &   V  &  BS97       \\ 
  48407.372  & 0.001   &   0.0043 &    0.0090 &   PE &  B98(1991)  \\ 
  48438.3944 & 0.0006  &   0.0009 & $-$0.0007 &   PE &  -          \\ 
  49501.7627 & 0.0003  &   0.0004 &    0.0008 &   PE &  BS97       \\ 
  52745.6521 & 0.0012  &   0.0017 &    0.0007 &   PE &  MK04       \\ 
  52758.0641 &         &   0.0035 & $-$0.0001 &   PE &  N04        \\ 
  53006.3222 &         &   0.0035 &    0.0011 &   PE &  N05        \\ 
  53124.2389 &         &   0.0035 & $-$0.0042 &   PE &  -          \\ 
  53461.4575*& 0.0004  &          & $-$0.0011 &   PE &  H06        \\ 
  53463.5256*& 0.0013  &          & $-$0.0018 &   PE &  D05        \\ 
\hline
\end{longtable}
\begin{longtable}{lllrlc}
\caption[]{\label{tab:adboo_tmins}
Times of $secondary$ minima for AD\,Boo.
See Table~\ref{tab:adboo_tminp} for details.}\\
\hline\hline\noalign{\smallskip}
HJD          & rms     &  rms     &   O-C     & Type &  Reference  \\
$-$ 2\,400\,000& publ. &  adopt.  & days      &      &             \\
\noalign{\smallskip}
\endfirsthead
\caption{continued.}\\
\hline\hline\noalign{\smallskip}
HJD          & rms     &  rms     &   O-C     &Observing&  Reference  \\
- 2\,400\,000& publ.   &  adopt.  & days      &method   &             \\
\hline\noalign{\smallskip}
\endhead
\hline
\noalign{\smallskip}
  15437.897  &         &   0.0320 &    0.0774 &   PG &  S68        \\ 
  18439.694  &         &   0.0320 &    0.0354 &   PG &  -          \\ 
  18799.671  &         &   0.0320 &    0.0400 &   PG &  -          \\ 
  19883.703  &         &   0.0320 &    0.0171 &   PG &  -          \\ 
  20636.749  &         &   0.0320 &    0.0173 &   PG &  -          \\ 
  20940.860  &         &   0.0320 &    0.0137 &   PG &  -          \\ 
  20967.771  &         &   0.0320 &    0.0302 &   PG &  -          \\ 
  21716.692  &         &   0.0320 &    0.0430 &   PG &  -          \\ 
  24995.683  &         &   0.0320 & $-$0.0251 &   PG &  -          \\ 
  25738.438  &         &   0.0320 &    0.0282 &   PG &  S63        \\ 
  25827.366  &         &   0.0320 & $-$0.0025 &   PG &  S63        \\ 
  26830.727  &         &   0.0320 & $-$0.0130 &   PG &  S68        \\ 
  27182.494  &         &   0.0320 &    0.0568 &   PG &  S63        \\ 
  27213.451  &         &   0.0320 & $-$0.0183 &   PG &  -          \\ 
  27573.458  &         &   0.0320 &    0.0163 &   PG &  -          \\ 
  27602.416  &         &   0.0320 &    0.0110 &   PG &  -          \\ 
  28996.735  &         &   0.0320 & $-$0.0459 &   PG &  S68        \\ 
  29495.393  &         &   0.0320 &    0.0296 &   PG &  S63        \\ 
  37017.537  &         &   0.0320 & $-$0.0088 &   PG &  -          \\ 
  40745.534  &         &   0.0150 & $-$0.0021 &   V  &  O119(1970),VB74 \\ 
  41041.378  &         &   0.0150 &    0.0025 &   V  &  O124(1971) \\ 
  41041.382  &         &   0.0150 &    0.0065 &   V  &  -          \\ 
  41134.446  &         &   0.0150 & $-$0.0259 &   V  &  VB74       \\ 
  41405.466  &         &   0.0150 & $-$0.0196 &   V  &  B2(1972),VB74 \\ 
  41434.442  &         &   0.0150 & $-$0.0069 &   V  &  B3(1972),VB74 \\ 
  41434.446  &         &   0.0150 & $-$0.0029 &   V  &  B3(1972)   \\ 
  41436.516  &         &   0.0150 & $-$0.0017 &   V  &  B4(1972)   \\ 
  41494.440  &         &   0.0150 & $-$0.0043 &   V  &  B3(1972),VB74 \\ 
  41763.339  &         &   0.0150 & $-$0.0502 &   V  &  B8(1973),VB74 \\ 
  41763.381  &         &   0.0150 & $-$0.0082 &   V  &  -          \\ 
  41794.399  &         &   0.0150 & $-$0.0223 &   V  &  B9(1973),VB74 \\ 
  41794.411  &         &   0.0150 & $-$0.0103 &   V  &  -          \\ 
  41912.365  &         &   0.0150 &    0.0217 &   V  &  B11(1973)  \\ 
  42156.453  &         &   0.0150 & $-$0.0095 &   V  &  B15(1974)  \\ 
  42158.525  &         &   0.0150 & $-$0.0063 &   V  &  -          \\ 
  42183.344  &         &   0.0150 & $-$0.0130 &   V  &  -          \\ 
  42183.369  &         &   0.0150 &    0.0120 &   V  &  -          \\ 
  42214.380  &         &   0.0150 & $-$0.0091 &   V  &  B16(1974)  \\ 
  42303.335  &         &   0.0150 & $-$0.0128 &   V  &  B17(1974)  \\ 
  42404.709  &         &   0.0150 & $-$0.0104 &   V  &  B19(1974)  \\ 
  42917.765  &         &   0.0150 & $-$0.0185 &   V  &  BS97       \\ 
  43190.826  &         &   0.0150 & $-$0.0401 &   V  &  -          \\ 
  43219.825  &         &   0.0150 & $-$0.0043 &   V  &  -          \\ 
  43689.457  &         &   0.0150 &    0.0085 &   V  &  B37(1978)  \\ 
  44334.900  &         &   0.0150 & $-$0.0163 &   V  &  BS97       \\ 
  44701.0941 & 0.0003  &   0.0010 & $-$0.0011 &   PE &  Z83        \\ 
  44730.0570 & 0.0005  &   0.0017 & $-$0.0015 &   PE &  -          \\ 
  44736.2642 & 0.0004  &   0.0014 & $-$0.0007 &   PE &  -          \\ 
  44779.686  &         &   0.0150 & $-$0.0239 &   V  &  BS97       \\ 
  45100.390  &         &   0.0150 &    0.0150 &   V  &  B60(1982)  \\ 
  45162.420  &         &   0.0150 & $-$0.0192 &   V  &  B61(1982)  \\ 
  46169.922  &         &   0.0150 & $-$0.0262 &   V  &  BS97       \\ 
  46217.514  &         &   0.0150 & $-$0.0168 &   V  &  I86        \\ 
  47580.8765 & 0.0005  &   0.0017 &    0.0019 &   PE & This paper  \\ 
  48360.8148 & 0.0002  &   0.0007 & $-$0.0001 &   PE &  -          \\ 
  49829.661  &         &   0.0035 & $-$0.0069 &   PE & BS97        \\ 
  50189.636  &         &   0.0035 & $-$0.0043 &   PE & -           \\ 
  52322.5851 & 0.0003  &   0.0031 &    0.0048 &   PE & B127(2002)  \\ 
  53522.4864*& 0.0020  &          & $-$0.0020 &   PE &  H06        \\ 

\hline
\end{longtable}
\clearpage
\twocolumn
\begin{table}
\caption[]{\label{tab:hwcma_tmin}Times of primary (P) minima of HW\,CMa.
O-C values are calculated for the
ephemeris given in Eq. \ref{eq:hwcma_eph}.
}
\begin{flushleft}
\begin{tabular}{llcrc} \hline
\hline\noalign{\smallskip}
HJD           & rms    & Type &   O-C      &  Reference  \\
$-$ 2\,400\,000 &      &      &            &              \\
\noalign{\smallskip}
\hline
\noalign{\smallskip}
48689.6471    &0.0004  & P    &   0.0000   & This paper \\
52279.6787    &0.0004  & P    &   0.0000   &     -      \\
\hline
\end{tabular}
\end{flushleft}
\end{table}
\begin{table}
\caption[]{\label{tab:swcma_tmin}Times of primary (P) and secondary (S)
minima of SW\,CMa.
O-C values are calculated for the
apsidal motion ephemeris given in Table \ref{tab:swcma_aps}.
F37 = Florja (\cite{florja37}).
H00 = Hipparcos 2000, unpublished.
LF94 = Lacy \& Fox (\cite{lacyfox94}).
W52 = Wenzel (\cite{wenzel52}); 
      photographic observations (rms of $0 \fd 01$ has been estimated).
Z65 = Ziegler (\cite{ziegler65});
      photographic observations (rms of $0 \fd 01$ has been estimated).
Times given in brackets have not been used for the ephemeris determination.
}
\begin{flushleft}
\begin{tabular}{llcrc} \hline
\hline\noalign{\smallskip}
HJD           & rms         & Type &   O-C      &  Reference  \\
$-$ 2\,400\,000 &           &      &            &              \\
\noalign{\smallskip}
\hline
\noalign{\smallskip}
25646.489     &0.010     & P    & $-0.087$   & W52 \\
25969.426     &0.010     & P    & $-0.094$   & - \\ 
25969.469     &0.010     & P    & $-0.051$   & - \\
26393.319     &0.010     & P    & $-0.064$   & - \\
26393.404     &0.010     & P    &   0.021    & - \\  
27160.327     &0.010     & P    & $-0.047$   & - \\
28250.320     &0.010     & P    &   0.012    & - \\
37696.372     &0.010     & P    & $-0.032$   & Z65 \\
38372.541     &0.010     & P    & $-0.026$   & - \\
38463.330     &0.010     & P    & $-0.065$   & - \\
46829.6482    &0.0003    & P    & $-0.0008$  & This paper \\
46839.7405    &0.0008    & P    & $-0.0005$  & - \\   
48272.8031    &0.0003    & P    &   0.0005   & - \\
49352.6453    &0.0004    & P    &   0.0005   & LF94 \\
52309.5963    &0.0002    & P    &   0.0005   & This paper \\
54075.6924    &0.0002    & P    & $-0.0005$  & - \\
25619.400     &0.010     & S    & $-0.035$   & W52 \\
26628.566     &0.010     & S    & $-0.067$   & F37   \\    
26709.324     &0.010     & S    & $-0.044$   & F37 \\  
27123.080     &0.010     & S    & $-0.060$   & F37   \\    
(27133.308)   &0.010     & S    &   0.077    & W52 \\    
(27780.015)   &0.010     & ?    &   0.897    & F37 \\   
27890.090     &0.010     & S    & $-0.040$   & -     \\    
28162.530     &0.010     & S    & $-0.083$   & W52 \\
28495.601     &0.010     & S    & $-0.047$   & - \\
30433.316     &0.010     & S    &   0.008    & - \\
34066.378     &0.010     & S    & $-0.042$   & - \\
36599.438     &0.010     & S    & $-0.068$   & Z65 \\
(37366.328)   &0.010     & S    & $-0.168$   & - \\     
46832.7710    &0.0010    & S    &   0.0001   & This paper \\
47579.5765    &0.0004    & S    & $-0.0007$  & - \\
47589.6690    &0.0004    & S    & $-0.0002$  & - \\
48164.905     & 0.003    & S    & $-0.007$   & H00       \\  
48265.8327    &0.0010    & S    &   0.0011   & This paper \\
49345.6754    &0.0006    & S    &   0.0022   & LF94 \\
52302.6227    &0.0002    & S    &   0.0001   & This paper \\
53644.8536    &0.0007    & S    & $-0.0020$  & - \\
\hline
\end{tabular}
\end{flushleft}
\end{table}
\begin{table}
\caption[]{\label{tab:v636cen_tmin}
Times of primary (P) and secondary (S) minima for V636\,Cen. 
O-C values are calculated from the apsidal motion parameters
given in Table~\ref{tab:v636cen_aps}.
H58 = Hoffmeister (\cite{h58}); rms of $0 \fd 03$ has been estimated.
H00 = Hipparcos 2000, unpublished.
}
\begin{flushleft}
\begin{tabular}{llcrc} \hline
\hline\noalign{\smallskip}
HJD           & rms         & Type &   O-C      &  Reference  \\
$-$ 2\,400\,000 &           &      &            &              \\
\noalign{\smallskip}
\hline
\noalign{\smallskip}
27977.317     & 0.030     & P    & $-0.010$   & H58          \\
28658.405     & 0.030     & P    & $-0.069$   & -            \\
28688.414     & 0.030     & P    & $-0.048$   & -            \\
34480.390     & 0.030     & P    &   0.037    & -             \\
34510.323     & 0.030     & P    & $-0.018$   & -             \\
34540.340     & 0.030     & P    &   0.012    & -             \\
34570.300     & 0.030     & P    & $-0.016$   & -             \\
46873.8014    &0.0001     & P    &   0.0003   & This paper    \\
47220.8003    &0.0001     & P    & $-0.0002$  &       -       \\
47250.7883    &0.0002     & P    &   0.0002   &       -       \\
47627.7751    &0.0001     & P    &   0.0000   &       -       \\
48167.538     &0.014      & P    & $-0.014$   & H00            \\
48364.6132    &0.0001     & P    & $-0.0002$  & This paper    \\
52305.84068   &0.00020    & P    & $-0.00056$ &       -       \\
52365.81615   &0.00015    & P    & $-0.00030$ &       -       \\
52682.82827   &0.00015    & P    &   0.00001  &       -       \\
54293.59119   &0.00010    & P    &   0.00023  &       -        \\
46871.7348    &0.0003     & S    & $-0.0002$  &       -       \\
47235.8713    &0.0005     & S    &   0.0007   &       -       \\
47248.7222    &0.0003     & S    & $-0.0003$  &       -       \\
52303.78204   &0.00030    & S    &   0.00051  &       -       \\
52436.58404   &0.00015    & S    &   0.00011  &       -       \\
52680.76871   &0.00040    & S    & $-0.00028$ &       -       \\
52710.75656   &0.00030    & S    & $-0.00007$ &       -       \\
53854.57063   &0.00030    & S    & $-0.00021$ &       -       \\
54291.53262   &0.00050    & S    & $-0.00096$ &       -        \\
\hline
\end{tabular}
\end{flushleft}
\end{table}
\begin{table*}
\caption[]{\label{tab:vzhya_tmin}
Times of primary (P) and secondary (S) minima for VZ\,Hya.
O-C values are calculated for the
ephemeris given in Eq.  \ref{eq:vzhya_eph} adopting a circular orbit.
References are:
B = BBSAG Bulletin, followed by volume number and year of publication.
BA74 = Baldwin (\cite{ba74}).
BAV = B.A.V. Mitt., followed by volume number and year of publication.
G53 = epoch given by Gaposchkin (\cite{gap53}).
L49 = Lause (\cite{lause49}).
O32 = epoch given by O'Connell (\cite{oconnell32}).
PS75 = Padalia \& Srivastava (\cite{ps75}). 
SC07 = Smith \& Caton (\cite{sc07}).
W46 = Wood (\cite{wood46}).
W70 = redetermined from  $B$ and $V$ data by Walker (\cite{walker70}).
W82 = Wolf et al. (\cite{wolf82}).
Observing methods are: PG photographic; V visual; PE photoelectric or CCD.
Times marked with an asterisk were published after we completed the
ephemeris calculations; including them makes no significant changes.
}
\begin{center}
\begin{tabular}{lllrllc} \hline
\hline\noalign{\smallskip}
HJD          & rms     &  rms     &   O-C     & Type& Observing& Reference  \\
$-$ 2\,400\,000& publ. &  adopt.  & days      &     &method&            \\
\noalign{\smallskip}
\hline
\noalign{\smallskip}
  21925.8246 & 0.0008  &   0.0011 &    0.0018 &  P  &  PG & O32 ($T_0$) \\ 
  26674.3410 & 0.0010  &   0.0120 &  $-0.0127$&  P  &  PG & G53 ($T_0$) \\ 
  27856.412  &         &   0.0063 &    0.0081 &  P  &  V  & L49         \\ 
  29604.794  &         &   0.0063 &    0.0014 &  P  &  V  & W46         \\ 
  29700.634  &         &   0.0063 &  $-0.0006$&  P  &  V  & -           \\ 
  30034.626  &         &   0.0063 &  $-0.0031$&  P  &  V  & -           \\ 
  39926.673  &         &   0.0063 &  $-0.0027$&  P  &  V  & BA74        \\ 
  40254.8602 & 0.0005  &   0.0007 &  $-0.0014$&  P  &  PE & W70         \\ 
  40998.362  &         &   0.0028 &  $-0.0004$&  P  &  PE & PS75        \\ 
  41033.212  &         &   0.0028 &  $-0.0020$&  P  &  PE & -           \\ 
  42354.668  &         &   0.0063 &  $-0.0026$&  P  &  V  & B18(1974)   \\ 
  42848.401  &         &   0.0063 &  $-0.0007$&  P  &  V  & B27(1976)   \\ 
  43577.372  &         &   0.0063 &  $-0.0090$&  P  &  V  & B37(1978)   \\ 
  44236.6532 &         &   0.0028 &  $-0.0040$&  P  &  PE & W82         \\ 
  47971.5876*& 0.0003  &          &    0.0003 &  P  &  PE & SC07        \\ 
  48267.8259 & 0.0001  &   0.0001 &  $-0.0001$&  P  &  PE & This paper  \\ 
  48270.7303 & 0.0001  &   0.0001 &    0.0000 &  P  &  PE & -           \\ 
  48273.6346 & 0.0001  &   0.0001 &    0.0001 &  P  &  PE & -           \\ 
  51253.4340 &         &   0.0063 &  $-0.0125$&  P  &  V  & BAV122(1999)\\ 
  51256.3520 &         &   0.0063 &    0.0012 &  P  &  V  & -           \\ 
  52702.6936*& 0.0001  &          &    0.0012 &  P  &  PE & SC07        \\ 
  27840.437  &         &   0.0063 &    0.0067 &  S  &  V  & L49         \\ 
  29681.751  &         &   0.0063 &  $-0.0056$&  S  &  V  & W46         \\ 
  29748.556  &         &   0.0063 &    0.0005 &  S  &  V  & -           \\ 
  30015.742  &         &   0.0063 &  $-0.0091$&  S  &  V  & -           \\ 
  40305.6848 & 0.0005  &   0.0015 &  $-0.0020$&  S  &  PE & W70         \\ 
  40654.201  &         &   0.0028 &  $-0.0019$&  S  &  PE & PS75        \\ 
  40686.154  &         &   0.0028 &    0.0038 &  S  &  PE & -           \\ 
  41743.305  &         &   0.0063 &  $-0.0104$&  S  &  V  & B8(1973)    \\ 
  48684.5934 & 0.0002  &   0.0006 &    0.0004 &  S  &  PE & This paper  \\ 
\hline
\end{tabular}
\end{center}
\end{table*}
\begin{table*}
\caption[]{\label{tab:wzoph_tminp}
Times of $primary$ minima for WZ Oph.
O-C  values are calculated for the
ephemeris given in Eq.  \ref{eq:wzoph_eph} assuming a circular orbit.
References are:
B = BBSAG Bulletin, followed by volume number and year of publication.
BAA = BAA VSS Circ., followed by volume number and year of publication.
BAV = B.A.V. Mitt., followed by volume number and year of publication.
BR = Brno Contr., followed by volume number and year of publication.
G38 = Gaposchkin (\cite{gap38}).
D77 = Dworak (\cite{d77}).
H01 = Hipparcos 2001, unpublished.
H06 = H\"ubscher et al. (\cite{h06}).
L21 = Leiner (\cite{leiner21}).
L26 = Leiner (\cite{leiner26}).
L36 = Lause (\cite{lause36}).
L38 = Lause (\cite{lause38}).
M29 = McLaughlin (\cite{mcl29}.
N05 = Nagai (\cite{n05}).
O = Orion (Diethelm \& Locker \cite{dl70b},\cite{dl71c}),
followed by volume number and year of publication.
P36 = Prager (\cite{gul36}).
P65 = Popper (\cite{dmp65}); determined by the present authors.
SC07 = Smith \& Caton (\cite{sc07}).
Observing methods are: PG photographic; V visual; PE photoelectric or CCD.
Times marked with an asterisk were published after we completed the
ephemeris calculations; including them makes no significant changes.
}
\begin{center}
\begin{tabular}{lllrlc} \hline
\hline\noalign{\smallskip}
HJD          & rms     &  rms     &   O-C     &Observing& Reference  \\
$-$ 2\,400\,000& publ. &  adopt.  & days      &method   &            \\
\noalign{\smallskip}
\hline
\noalign{\smallskip}
  14766.779  &         &   0.0340 &  $-0.0211$ &   PG & G38         \\ 
  16402.534  &         &   0.0340 &  $-0.0172$ &   PG & -           \\ 
  16678.62   &         &   0.0340 &  $-0.0427$ &   PG & -           \\ 
  16975.67   &         &   0.0340 &  $-0.0217$ &   PG & -           \\ 
  17707.813  &         &   0.0340 &    0.0076  &   PG & -           \\ 
  17728.754  &         &   0.0340 &    0.0311  &   PG & -           \\ 
  18523.56   &         &   0.0340 &  $-0.0292$ &   PG & -           \\ 
  19531.77   &         &   0.0340 &  $-0.0443$ &   PG & -           \\ 
  20778.51   &         &   0.0340 &    0.0106  &   PG & -           \\ 
  22531.385  &         &   0.0110 &  $-0.0037$ &   V  &  L21        \\ 
  22836.74   &         &   0.0340 &  $-0.0447$ &   PG &  G38        \\ 
  22941.379  &         &   0.0110 &    0.0066  &   V  &  L21        \\ 
  22962.303  &         &   0.0110 &    0.0131  &   V  &  -          \\ 
  23284.431  &         &   0.0110 &    0.0111  &   V  &  L26        \\ 
  23535.440  &         &   0.0110 &    0.0097  &   V  &  -          \\ 
  23581.451  &         &   0.0110 &    0.0021  &   V  &  -          \\ 
  23648.393  &         &   0.0110 &    0.0080  &   V  &  -          \\ 
  24355.405  &         &   0.0110 &    0.0073  &   V  &  -          \\ 
  24702.626  &         &   0.0340 &  $-0.0027$ &   PG &  G38        \\ 
  24744.520  &         &   0.0340 &    0.0562  &   PG &  -          \\ 
  25363.601  &         &   0.0340 &  $-0.0218$ &   PG &  -          \\ 
  25384.477  &         &   0.0340 &  $-0.0634$ &   PG &  -          \\ 
  25384.563  &         &   0.0340 &    0.0226  &   PG &  -          \\ 
  25388.729  &         &   0.0110 &    0.0051  &   V  &  M29        \\ 
  27250.410  & 0.004   &   0.0180 &    0.0256  &   V  &  D77        \\ 
  27639.444  &         &   0.0110 &  $-0.0065$ &   V  &  L36        \\ 
  27660.376  &         &   0.0110 &    0.0079  &   V  &  -          \\ 
  27890.448  &         &   0.0110 &  $-0.0129$ &   V  &  -          \\ 
  27961.570  &         &   0.0110 &  $-0.0106$ &   V  &  -          \\ 
  28003.411  &         &   0.0110 &  $-0.0046$ &   V  &  -          \\ 
  28455.228  &         &   0.0110 &  $-0.0064$ &   V  &  L38        \\ 
  28555.627  &         &   0.0110 &  $-0.0115$ &   V  &  -          \\ 
  28664.418  &         &   0.0110 &    0.0083  &   V  &  -          \\ 
  28689.498  &         &   0.0110 &  $-0.0127$ &   V  &  -          \\ 
  28819.172  &         &   0.0110 &  $-0.0274$ &   V  &  -          \\ 
  28844.310  &         &   0.0110 &    0.0095  &   V  &  -          \\ 
  35650.8670 & 0.0010  &   0.0028 &    0.0009  &   PE &  P65        \\ 
  40859.331  &         &   0.0110 &  $-0.0011$ &   V  &  O121(1970) \\ 
  41135.427  &         &   0.0110 &  $-0.0165$ &   V  &  O126(1971) \\ 
  46243.508  &         &   0.0110 &    0.0027  &   V  &  B77(1985)  \\ 
  47381.420  &         &   0.0110 &    0.0008  &   V  &  BR30(1992) \\ 
  47724.471  &         &   0.0110 &    0.0043  &   V  &  BAV56(1989)\\ 
  48004.7536*& 0.0002  &          &  $-0.0081$ &  PE  &  SC07       \\ 
  48042.420  &         &   0.0110 &    0.0068  &   V  &  B95(1990)  \\ 
  48088.453  &         &   0.0110 &    0.0212  &   V  &  B96(1990)  \\ 
  48372.9097 & 0.0004  &   0.0011 &  $-0.0006$ &   PE &  This paper \\ 
  48452.3967 &         &   0.0003 &  $-0.0002$ &   PE &  BAA91(1997)\\ 
  48795.445  & 0.010   &   0.0280 &    0.0005  &   PE &  B102(1992) \\ 
  48883.3011 & 0.0008  &   0.0022 &    0.0030  &   PE &  B103(1993) \\ 
  50535.7849 & 0.0003  &   0.0008 &    0.0016  &   PE &  This paper \\ 
  52054.396  &         &   0.0110 &  $-0.0003$ &   V  &  BAV143(2001)\\
  52100.411  &         &   0.0110 &  $-0.0039$ &   V  &  BAV154(2002)\\
  53476.7886*& 0.0001  &          &    0.0000  &  PE  &  SC07       \\ 
\hline
\end{tabular}
\end{center}
\end{table*}
\begin{table*}
\caption[]{\label{tab:wzoph_tmins}
Times of $secondary$ minima for WZ Oph.
See Table~\ref{tab:wzoph_tminp} for details.
}
\begin{center}
\begin{tabular}{lllrrc} \hline
\hline\noalign{\smallskip}
HJD          & rms     &  rms     &   O-C     &Observing& Reference  \\
$-$ 2\,400\,000& publ. &  adopt.  & days      &method   &            \\
\noalign{\smallskip}
\hline
\noalign{\smallskip}
  16379.51   &         &   0.0340 &  $-0.0319$ &   PG &  G38        \\ 
  16634.72   &         &   0.0340 &  $-0.0159$ &   PG &  -          \\ 
  17107.531  &         &   0.0340 &    0.0589  &   PG &  -          \\ 
  17341.73   &         &   0.0340 &  $-0.0185$ &   PG &  -          \\ 
  18027.80   &         &   0.0340 &  $-0.0436$ &   PG &  -          \\ 
  19596.592  &         &   0.0340 &  $-0.0667$ &   PG &  -          \\ 
  19596.634  &         &   0.0340 &  $-0.0247$ &   PG &  -          \\ 
  20604.847  &         &   0.0340 &  $-0.0368$ &   PG &  -          \\ 
  21675.81   &         &   0.0340 &  $-0.0516$ &   PG &  -          \\ 
  22215.56   &         &   0.0340 &    0.0261  &   PG &  -          \\ 
  22487.458  &         &   0.0110 &  $-0.0039$ &   V  &  L21        \\ 
  22897.4557 &         &   0.0340 &    0.0101  &   PG &  P36        \\ 
  22897.458  &         &   0.0110 &    0.0124  &   V  &  L21        \\ 
  22918.374  &         &   0.0110 &    0.0109  &   V  &  -          \\ 
  23516.615  &         &   0.0110 &    0.0104  &   V  &  L26        \\ 
  23692.315  &         &   0.0110 &    0.0031  &   V  &  -          \\ 
  23859.662  &         &   0.0110 &    0.0099  &   V  &  -          \\ 
  24273.807  &         &   0.0340 &  $-0.0123$ &   PG &  G38        \\ 
  24286.376  &         &   0.0110 &    0.0062  &   V  &  L26        \\ 
  25127.259  & 0.005   &   0.0225 &    0.0043  &   V  &  D77        \\ 
  25361.511  &         &   0.0340 &  $-0.0201$ &   PG &  G38        \\ 
  26482.74   &         &   0.0340 &    0.0291  &   PG &  G38        \\ 
  26545.487  & 0.006   &   0.0270 &    0.0235  &   V  &  D77        \\ 
  27570.425  &         &   0.0110 &    0.0023  &   V  &  L36        \\ 
  27637.360  &         &   0.0110 &    0.0012  &   V  &  -          \\ 
  27980.411  &         &   0.0110 &    0.0047  &   V  &  -          \\ 
  27984.593  &         &   0.0110 &    0.0032  &   V  &  -          \\ 
  28026.418  &         &   0.0340 &  $-0.0069$ &   PG &  -          \\ 
  28068.266  &         &   0.0110 &    0.0060  &   V  &  -          \\ 
  28072.420  &         &   0.0110 &  $-0.0235$ &   V  &  -          \\ 
  28298.356  &         &   0.0110 &    0.0031  &   V  &  L38        \\ 
  28390.384  &         &   0.0110 &  $-0.0060$ &   V  &  -          \\ 
  28457.325  &         &   0.0110 &  $-0.0011$ &   V  &  -          \\ 
  28687.434  &         &   0.0110 &    0.0150  &   V  &  -          \\ 
  28754.337  &         &   0.0110 &  $-0.0181$ &   V  &  -          \\ 
  28779.454  &         &   0.0110 &  $-0.0021$ &   V  &  -          \\ 
  33063.382  & 0.002   &   0.0090 &    0.0149  &   V  &  D77        \\ 
  35648.7748 & 0.0007  &   0.0014 &    0.0005  &   PE &  P65        \\ 
  40836.327  &         &   0.0110 &    0.0042  &   V  &  O121(1970) \\ 
  41522.387  &         &   0.0110 &  $-0.0309$ &   V  &  B4(1972)   \\ 
  41522.403  &         &   0.0110 &  $-0.0149$ &   V  &  -          \\ 
  41522.408  &         &   0.0110 &  $-0.0099$ &   V  &  -          \\ 
  41819.460  &         &   0.0110 &    0.0131  &   V  &  B9(1973)   \\ 
  42183.416  &         &   0.0110 &    0.0040  &   V  &  B15(1974)  \\ 
  43689.479  &         &   0.0110 &    0.0046  &   V  &  B37(1978)  \\ 
  46241.431  &         &   0.0110 &    0.0174  &   V  &  B77(1985)  \\ 
  48370.8180 & 0.0003  &   0.0006 &  $-0.0005$ &   PE &  This paper \\ 
  48408.468  & 0.006   &   0.0120 &  $-0.0021$ &   PE &  B98(1991)  \\ 
  48521.431  & 0.008   &   0.0160 &    0.0062  &   PE &  H01        \\ 
  49115.4800 & 0.0007  &   0.0014 &  $-0.0027$ &   PE &  BAV62(1993)\\ 
  49115.4824 & 0.0007  &   0.0014 &  $-0.0003$ &   PE &  -          \\ 
  49504.5461 & 0.0012  &   0.0024 &  $-0.0028$ &   PE &  B107(1994) \\ 
  49529.6504 & 0.0002  &   0.0004 &    0.0005  &   PE &  This paper \\ 
  53098.180  &         &   0.0110 &  $-0.0012$ &   V  &  N05        \\ 
  53901.4151*& 0.0002  &          &    0.0006  &   PE &  H06        \\ 
\hline
\end{tabular}
\end{center}
\end{table*}

\end{appendix}



\begin{thebibliography}{}
\bibitem[1991]{ja91}                                   
Andersen, J. 1991, \aapr, 3, 91
\bibitem[1980]{jaetal80}                               
Andersen, J., Clausen, J.~V., \& Nordstr\"om, B. 1980,
in Close Binary Stars: Observations and 
Interpretation (IAU Symp. No. 88), ed. M. J. Plavec,
D.~M. Popper, and R.~K. Ulrich. Reidel, Dordrecht, p. 81
\bibitem[1974]{ba74}                                   
Baldwin, M.~E., 1974, J. American Ass. Var. Star Obs. 3, 60
\bibitem[1978]{ba78}                                   
Baldwin, M.~E., 1978, J. American Ass. Var. Star Obs. 7, 28
\bibitem[1997]{bs97}                                   
Baldwin, M.~E., \& Samolyk, G. 1997,
Obs. minima timings of ecl. bin. No. 4
\bibitem[1978]{cester78}                               
Cester, B., Fedel, B., Giuricin, G., et al. 1978,
\aaps, 32, 351
\bibitem[2007]{claret07}                               
Claret, A. 2007, \aap, 475, 1019
\bibitem[1999a]{granada99a}
Clausen, J.~V., Helt, B.~E., \& Olsen, E.~H. 1999a,
in Theory and Tests of Convection in Stellar
Structure, ed. A. Gim\'enez, E.~F. Guinan, \& B. Montesinos, 
ASP Conference Series, 173, 321
\bibitem[1999b]{granada99b}
Clausen, J.~V., Baraffe, I., Claret, A., \&
VandenBerg, D.~B. 1999b, 
in Theory and Tests of Convection in Stellar
Structure, ed. A. Gim\'enez, E.~F. Guinan, \& B. Montesinos, 
ASP Conference Series, 173, 265
\bibitem[2001]{jvcetal01}                              
Clausen, J.~V., Helt, B.~E., \& Olsen, E.~H. 2001,
\aap, 374, 980
\bibitem[2008]{avw08}                                   
Clausen, J.~V., Torres, G., Bruntt, H., et al. 2008,
\aap, in press
\bibitem[2005]{d05}                                     
Diethelm, R. 2005,
Inf. Bull. Var. Stars, 5653
\bibitem[1970a]{dl70a}                                   
Diethelm, R., \& Locher, K. 1970a, Orion No. 119, 126
\bibitem[1970b]{dl70b}                                   
Diethelm, R., \& Locher, K. 1970b, Orion No. 121, 191
\bibitem[1971a]{dl71a}                                   
Diethelm, R., \& Locher, K. 1971a Orion No. 124, 91 
\bibitem[1971b]{dl71b}                                   
Diethelm, R., \& Locher, K. 1971b, Orion No. 125, 111
\bibitem[1971c]{dl71c}                                   
Diethelm, R., \& Locher, K. 1971c, Orion No. 126, 142
\bibitem[1973]{d73}                                    
Dworak, T.~Z. 1973, Inf. Bull. Var. Stars, 846 
\bibitem[1977]{d77}                                    
Dworak, T.~Z. 1977, Acta Astron., 27, 151
\bibitem[1989]{do89}                                   
Dworak, T.~Z., \& Oblak, E. 1989, 
Inf. Bull. Var. Stars, 3399
\bibitem[1997]{hip97}
ESA 1997, The Hipparcos and Tycho Catalogues, 
ESA SP-1200
\bibitem[1937]{florja37}                               
Florja, N. 1937, 
Publ. Sternberg State Astr. Inst. 8, No. 2
\bibitem[1938]{gap38}                                  
Gaposchkin, S 1938, Harvard Bull. No. 907
\bibitem[1953]{gap53}                                  
Gaposchkin, S. 1953, Harvard Annals 113, 69
\bibitem[1994]{agc94}                                  
Gim\'enez, A. 1994, Experimental Astr. 5, 91
\bibitem[1983]{ggp83}                                  
Gim\'enez, A., \& Garcia-Pelayo, J.
1983, \apss, 92, 203
\bibitem[1995]{gb95}                                   
Gim\'enez, A., \& Bastero, M. 1995, \apss, 226, 99
\bibitem[1975]{hh75}                                   
Hilditch, R.~W., \& Hill, G. 1975, 
\mnras, 79, 101
\bibitem[1932]{h32}                                    
Hoffmeister, C. 1932, Astron. Nachr., 242, 131
\bibitem[1958]{h58}
Hoffmeister, C. 1958,                                  
Ver\"offentlichungen der Sternwarte in Sonneberg, 
3, 439
\bibitem[1988]{hsm88}
Houk, N. \& Smith-Moore, M., 1988,                     
Michigan Spectral Catalogue, Vol. 4,
Univ. of Michigan, Ann Arbor
\bibitem[2006]{h06}                                    
H\"ubscher, J., Paschke, A., \& Walter, F. 2006,
Inf. Bull. Var. Stars, 5731
\bibitem[1986]{i86}                                    
Isles, J. 1986, British Astron. Ass. Var. Star Sect.
Circ., 63, 19
\bibitem[1996]{jordi96}                                
Jordi, C., Figueras, F., Torra, J., Asiain, R.
1996, \aaps, 115, 401
\bibitem[1974]{koch74}                                 
Koch, R.~H. 1974, \aj, 79, 34
\bibitem[2001]{kreiner01}                              
Kreiner, J.~M., Kim, C.~H., \& Nha, I.~S. 2001, An Atlas of O$-$C
Diagrams of Eclipsing Binary Stars (Krakow: Wydawnictwo Naukowe Akad.\
Pedagogicznej)
\bibitem[1956]{kvw56}                                  
Kwee, K.~K., \& van~Woerden, H. 1956, \bain, 12, 327
\bibitem[1977]{lacy77}                                 
Lacy, C.~H. 1977, \apjs, 34, 479
\bibitem[1984]{lacy84}                                 
Lacy, C.~H. 1984, Inf. Bull. Var. Stars, 2489
\bibitem[1985]{lacy85}                                 
Lacy, C.~H. 1985, Inf. Bull. Var. Stars, 2685
\bibitem[1992]{lacy92}                                 
Lacy, C.~H. 1992, \aj, 104, 801 
\bibitem[1997a]{lacy97a}                               
Lacy, C.~H. S. 1997a, \aj, 113, 1406
\bibitem[1997b]{lacy97b}                               
Lacy, C.~H. S. 1997b, \aj, 113, 2226
\bibitem[2002]{lacy02}                                 
Lacy, C.~H. S. 2002, \aj, 124, 1162
\bibitem[1994]{lacyfox94}                              
Lacy, C.~H.~S., \& Fox, G.~W. 1994,
Inf. Bull. Var. Stars, 4009 
\bibitem[1965]{lk65}                                   
Lafler, J., \& Kinman, T.~D. 1965, \apjs, 11, 216
\bibitem[1998]{al98}                                   
Larsen, A. 1998, 
Master Thesis, Copenhagen University
\bibitem[1936]{lause36}                                
Lause, F. 1936, Astron. Nachr., 259, 189
\bibitem[1938]{lause38}                                
Lause, F. 1938, Astron. Nachr., 266, 17 
\bibitem[1949]{lause49}                                
Lause, F. 1949, Astron. Nachr., 277, 40         
\bibitem[1921]{leiner21}                               
Leiner, E. 1921, Astron. Nachr., 216, 23
\bibitem[1926]{leiner26}                               
Leiner, E. 1926, Astron. Nachr., 226, 179
\bibitem[1992]{liu92}                                  
Liu, O., Andersen, J., Clausen, J.~V., Nordstr\"om, B., 
\& Stefanik, R. P. 1992, Inf. Bull. Var. Stars, 3813
\bibitem[1929]{mcl29}                                  
McLaughlin, D.~B. 1929, \aj, 39, 87
\bibitem[2004]{mk04}                                   
Maciejewski, G. \& Karska, A., 2004,
Inf. Bull. Var. Stars, 5494
\bibitem[2007]{mrj07}                                  
Morales, J.~C., Ribas, I., \& Jordi, C. 2007,
astro-ph/07113523
\bibitem[2004]{n04}                                    
Nagai, K., 2004, VSOLJ Var. Star Bull. 42
\bibitem[2005]{n05}                                    
Nagai, K., 2005, VSOLJ Var. Star Bull. 43
\bibitem[1932]{oconnell32}                             
O'Connell, J.~K. 1932, Harvard Bull, No. 889, 7         %
\bibitem[1983]{olsen83}                                
Olsen, E.~H. 1983, \aaps, 54, 55
\bibitem[1988]{olsen88}                                
Olsen, E.~H. 1988, \aaps, 189, 173
\bibitem[1993]{olsen93}                                
Olsen, E.~H. 1993, \aaps, 102, 89
\bibitem[1994a]{olsen94a}                              
Olsen, E.~H. 1994a, \aaps, 104, 429
\bibitem[1994b]{olsen94b}                              
Olsen, E.~H. 1994b, \aaps, 106, 257
\bibitem[1984]{olsenperry84}                           
Olsen, E.~H., Perry, C.~L. 1984, \aaps, 56, 229
\bibitem[1986]{p86}                                    
Padalia, T.~D. 1986, \apss, 119, 395
\bibitem[1975]{ps75}                                   
Padalia, T.~D., \& Srivastava, R.~K. 1975, \apss, 35, 349
\bibitem[1985]{petal85}                                
Pastori, L., Pasinetti, E., Antonello, E., \&
Malaspina, G. 1985, in  Calibration of Fundamental
Stellar Quantities (IAU Symp. No. 111), ed. D. S. Hayes, 
L. E. Pasinetti, \& A. G. Davis Philip. 
Reidel, Dordrecht, p. 455
\bibitem[1969]{perry69}                                
Perry, C.~L., 1969, \aj, 74, 705
\bibitem[1917]{p17}                                    
Pickering, E.~C. 1917, Harvard Circ. No. 201
\bibitem[1965]{dmp65}                                  
Popper, D.~M. 1965, \apj, 141, 126
\bibitem[1966]{dmp66}                                  
Popper, D.~M. 1966, \aj, 71, 175
\bibitem[1976]{dmp76}                                  
Popper, D.~M. 1976, \apss, 45, 391
\bibitem[1980]{dmp80}                                  
Popper, D.~M. 1980, \araa, 18, 115
\bibitem[1993]{dmp93}                                  
Popper, D.~M. 1993, \apj, 404, L67
\bibitem[1996]{dmp96}                                  
Popper, D.~M. 1996, \apjs, 106, 133
\bibitem[1997]{dmp97}                                  
Popper, D.~M. 1997, \aj, 114, 1195
\bibitem[1998a]{dmp98a}                                
Popper, D.~M. 1998a, \aj, 115, 338
\bibitem[1998b]{dmp98b}                                
Popper, D.~M. 1998b, \pasp, 110, 919 
\bibitem[1981]{dmpetzel81}                             
Popper, D.~M., \& Etzel, P.B. 1981, \aj, 86, 102
\bibitem[1994]{dmpj94}                                 
Popper, D.~M., \& Jeong, Y.-C. 1994, \pasp, 106, 189
\bibitem[1970]{dmpetal70}                              
Popper, D.~M., J{\o}rgensen, H.~E., Morton, D.~C.,
\& Leckrone, D.~S. 1970, \apj, 161, L57
\bibitem[1936]{gul36}                                  
Prager, R. 1936, in
Geschichte und Literatur des Lichtwesels der 
Ver\"anderlichen Sterne, 2, II
\bibitem[2004]{ir04}                                   
Ribas, I. 2004, \nar, 48, 731
\bibitem[2006]{ir06a}                                   
Ribas, I. 2006, \apss, 304, 89
\bibitem[2006]{ir06b}                                   
Ribas, I. 2006, in Astrophysics of Variable Stars,
ed. C. Sterken, \& C. Aerts, 
ASP Conference Series 349, 55
\bibitem[2000a]{ir00a}                                   
Ribas, I., Jordi, C., Torra, J., \& Gim\'enez, A. 2000a,
\mnras, 313, 99
\bibitem[2000b]{ir00b}                                   
Ribas, I., Jordi, C., Torra, J., \& Gim\'enez, A. 2000b,
\mnras, 318, L55
\bibitem[1937]{s37}                                    
Sanford, R.~F. 1937, \apj, 86, 154
\bibitem[1980]{sb80}                                   
Shallis, M.~J., \& Blackwell, D.~E. 1980, \aap, 81, 336
\bibitem[2007]{sc07}                                   
Smith, A.~B. \& Caton, D.~B. 2007,
Inf. Bull. Var. Stars, 5745
\bibitem[1987]{s87}                                    
Srivastava, R.~K. 1987, \apss, 133, 71
\bibitem[1956]{sm56}                                   
Strohmeier, W., Kippenhahn, R., \& Geyer, E. 1956,
Kleine Ver. der Remeis-Sternwarte Bamberg, No. 15
\bibitem[1963]{sm63}                                   
Strohmeier, W., Knigge, R., \& Ott, H. 1963,
Ver. der Remeis-Sternwarte Bamberg, 5, No. 17
\bibitem[1968]{sm68}                                   
Strohmeier, W., \& Baurnfeind, H. 1968,
Ver. der Remeis-Sternwarte Bamberg, 7, 72
\bibitem[1945]{struve45}                               
Struve, O. 1945, \apj, 102, 74
\bibitem[2006]{wt06}                                   
Torres, G., Lacy, C.~H., Marschall, L.~A., 
Sheets, ,H.~A, \& Mader, J.~A. 2006, \aj, 640, 1018 
\bibitem[1974]{vb74}                                   
van Buren, D. 1974,
J. American Ass. Var. Star obs., 3, No. 1, 6
\bibitem[1970]{walker70}                               
Walker, R.~L. 1970, \aj, 75, 720
\bibitem[1952]{wenzel52}                               
Wenzel, W. 1952, Mitt. Ver. Sterne 2, 185
\bibitem[1983]{wk83}                                   
Wolf, G.~W., \& Kern, J.~T. 1983, \apjs, 52, 429
\bibitem[1982]{wolf82}                                 
Wolf, G.~W., Kern, J.~T., Hayes, T.~L., \& Chaffin, C.~R.
1982, Inf. Bull. Var. Stars, 2185
\bibitem[1971]{wood71}                                 
Wood, D.~B. 1971, \aj, 76, 701
\bibitem[1946]{wood46}                                 
Wood, F.~B. 1946, Princeton Contrib. 21, 1
\bibitem[1997]{wds}                                    
Worley, C.~E., \& Douglass, G.~G. 1997, \aaps, 125, 523
\bibitem[1989]{zz89}                                   
Zhai, D., \& Zhang, X. 1989, 
Publ. Beijing Astron. Obs., 12, 38
\bibitem[1982]{zhai82}                                 
Zhai, D.-S., Zhang, R.-X., \& Zhang, J.-T. 1982,
Acta Astrophys. Sinica, 2, 224
\bibitem[1983]{zhai83}                                 
Zhai, D.-S., Zhang, R.-X., \& Zhang, J.-T. 1983,
Inf. Bull. Variable Stars, 2275
\bibitem[1965]{ziegler65}                              
Ziegler, E. 1965, Mitt. Ver. Sterne, 2, 185
\end{thebibliography}
\end{document}